\documentclass{scrartcl}

\usepackage[utf8]{inputenc} 
\usepackage[T1]{fontenc}

\usepackage{amsmath} 
\usepackage{amssymb} 

\usepackage{graphicx} 
\usepackage[colorlinks,urlcolor=cyan,citecolor=blue,linkcolor=blue]{hyperref} 
\usepackage[parfill]{parskip}  
\usepackage{authblk}

\usepackage{longtable}
\usepackage{caption}
\usepackage{booktabs}

\usepackage{tikz}
\usetikzlibrary{positioning, arrows.meta}

\usepackage{natbib}

\author[1]{Théo Voldoire}
\author[2,3]{Robin J. Ryder}
\author[4]{Ryan Lahfa}
\affil[1]{Statistics Department, Harvard University, Cambridge, United States}
\affil[2]{Department of Mathematics, Imperial College, London, United Kingdom}
\affil[3]{CEREMADE, CNRS, Université Paris-Dauphine, Université PSL, Paris, France}
\affil[4]{DIENS, CNRS, \'Ecole normale sup\'erieure, Université PSL, Paris, France}
\date{}

\title{Some are observed, all leave traces: whole-population modeling of French elite civil servants' career paths}
\providecommand{\subtitle}[1]{}

\begin{document}
\maketitle

\section*{Abstract}

Elite civil servants may come and go between the public and private sectors throughout their career, a process of particular interest for the public and social scientists. 
However, data to document such processes are rarely completely available: we need inference tools that can account for many missing values. We consider public-private paths of elite French civil servants and introduce binary Markov switching models with Bayesian data augmentation. 
Our procedure relies on two complementary data sources: 
(1) detailed observations of some individual trajectories obtained from LinkedIn;  
(2) less informative ``traces'' left by all individuals in the administrative record, which we model for missing data imputation. 
This model class maintains the properties of hidden Markov models and enables a tailored sampler to target the posterior, yet allows for varying parameters across individuals and time. 
By integrating the two sources, we can consider the whole population rather than just a sample, and avoid the biases that would stem from using only a single source. 
We demonstrate this allows to properly test substantive hypotheses on career paths  across a variety of public organizations. We notably show that the probability for ENA graduates to exit the public sector has not increased since 1990, but that the probability they return has increased. We identify three clusters of organizations, with distinct patterns of public-private behaviors.\\

\textbf{Keywords:} Bayesian Markov Switching Model, Career Modeling, Public-Private Paths, Civil Servants, Missing Values, Data Augmentation

\section{Introduction}\label{section:introduction}

Many important questions in the social sciences require understanding career trajectories of certain populations. This is an important aspect when describing political \citep{jackle2018temporal}, economic \citep{dudouet_les_2010}, academic \citep{angermuller2023careers} and administrative \citep{rouban_norme_2014} elites, but also when studying specific professions like musicians \citep{abbott1990measuring}, journalists \citep{machut2023emploi}, or even athletes \citep{li2018journey}. Studying careers is seen as interesting in its own right, but also to describe broader social mechanisms such as lobbying and revolving doors \citep{blanes2012revolving} or the spatial structuring of finance \citep{buhlmann2024career}.

However, a key problem in career modeling is data missingness. It is very rare that all data relevant to the career of a certain population are available in a public record or can even be collected because of resource constraints, which is problematic for two reasons. First, for a given defined population, inference can be made more difficult and less accurate if a majority of individuals cannot be observed. This makes the testing of interesting interaction effects impossible in practice. Second, researchers often choose to only study populations that can be entirely observed. This forces them to study the groups that are the most visible, such as top political elites. Both these issues thus constrain the breadth of questions that can be addressed using career modeling. 

Our focus is the study of public-private paths of elite civil servants in France, which have already been the subject of many quantitative publications \citep{bouzidi_pantouflage_2010,rouban_norme_2014,kolopp2021pantoufler}. Public-private paths broadly understood for policy-relevant positions are sometimes described using the term ``revolving doors'', because these back-and-forth switches resemble a physical revolving door \citep{seabrooke2021revolving,france_sphere_2017}, but other scholars define revolving doors to be a much more precise and limited in scope phenomenon. Unassuming about this definitional debate, we denote this problem as the study of public-private paths of French elite civil servants, and acknowledge that our article falls in the first category of works.

Previous quantitative works of French civil servants have focused on data for a small number of individuals, typically collected by hand. Instead, our aim is to construct and apply a methodology to analyse the whole population, thanks to two complementary sources of data.
First, it is relatively easy to describe the career paths of a portion of elite civil servants, by collecting data from digital networking websites such as LinkedIn on which they are registered. However, a majority of civil servants do not have a LinkedIn profile, and further, it is not always the case that a profile is fully completed from the starting point of one's career. 
We thus complement the analysis with a second data source which covers the entire population over their entire public service career: to various degrees, civil servants leave  ``traces'' of their activity, when they sign legal documents, or obtain important promotions or bonuses which are recorded in administrative records published in the Journal Officiel de la République Française (JORF). 
We describe these two data sources in detail below. Neither source taken alone can lead to robust inference on career paths: the LinkedIn data alone is insufficient as it concerns only a non-random subpopulation (in particular, individuals who have never left the public sector are less likely to be present on LinkedIn); the JORF data alone is insufficient as it contains too little information on its own.

We showcase that by combining these two sources and modeling them jointly, it is possible to reconstruct trajectories of large civil servant populations and draw interpretations that contribute to the social scientific literature: the union of the two data sources is greater than the sum of its parts. In addition, our population is  more diverse and an order of magnitude larger compared to usual prosopography studies of the domain. Our methodology allows for analyses that are also targeted at less visible, larger professional groups. 
We describe our population of interest and data sources in Section \ref{section:setting}. 

In Section \ref{sec:model}, we build hidden Markov models with covariates and autoregressive components and describe our inference algorithm. 
Section \ref{ss:synthetic}
showcases, using plausible synthetic data examples, that these components are required for such a modeling task while remaining simple enough for substantive hypothesis testing. We then apply these methods to data about French civil servants from 1990 to 2022 in  Section \ref{section:applied}, demonstrating the usefulness of this strategy and expanding the social scientific literature of administrative elites in meaningful ways. 
We consider in particular differences in behavior between organizations and through time.

Our methodology could plausibly be adapted to other the career analysis of other professional groups with dual data sources: one with high information about a minority of the population (possibly a hand curated data base), the other with low information traces on the totality of the population.

%


\section{Problem setting} \label{section:setting}

\subsection{Population of interest}
\label{sss:popdef}

We consider elite French civil servants over the time frame 1990 to 2022. 
Top administrators are divided between groups and services that are sometimes tied to certain special statutes. While recent regulatory changes have attempted to increase mobility between them and foster a common organizational culture, they remain useful categories to base our approach upon.

Our study focuses on the following groups: three inspection groups whose mission is to audit public organizations organized thematically around financial affairs (igf, Inspection Générale des Finances), social affairs (igas, Inspection Générale des Affaires Sociales), and administration (iga, Inspection Générale de l'Administration); the Cour des Comptes (ccomptes), the supreme financial jurisdiction  with a similar financial auditing function; the Conseil d'État (ce), whose members serve both as judges in the Supreme Administrative Court and as legal advisors to the government; the \textit{préfets} (cprefet), state administration representatives  in regions and departments; ambassadors (cdiplo, Corps Diplomatique); administrators of the national statistical and economics office (insee) who can serve key roles in the definition of economic policies; and finally, directors and assistant directors of two powerful services of the Ministry of the Economy, treasury (dgtresor), and public finances (dgfip). Additionally, we include in our study all graduates of 
the National School of Administration (\'Ecole Nationale d'Administration, hereafter ENA). 
Our study includes all civil servants who joined or were promoted within one of these groups over the time frame 1990-2022.

These populations can be defined exhaustively, since all members of these groups are systematically described in the Journal Officiel de la République Française (JORF). This selection is voluntarily diverse and encompassing, as one of our goals is to describe how public-private mobility is a heterogeneous phenomenon. 
We obtained the complete list of population members from JORF, then used the LinkedIn API to find the profiles of these individuals. 
For each person in our population, we searched on LinkedIn for up to three profiles that share the same name, and then only kept the ones that list an experience in our starting set of organizations. For example, if an individual is included because they are part of the Statistical office, we only keep profiles with such an affiliation, using a rule-based approach. 
Overall, we designed our scheme to minimize false matchings between sources rather than maximizing the number of profiles that would be matched; manual inspection shows that it is very robust to homonymy problems in the matching between JORF and LinkedIn. 
The profiles output by the API queries were filtered using rules described in Appendix \ref{ap:data}, then all checked by hand. 

\begin{figure}
	\begin{minipage}{0.48\textwidth}
		\centering
		\includegraphics[width=\linewidth]{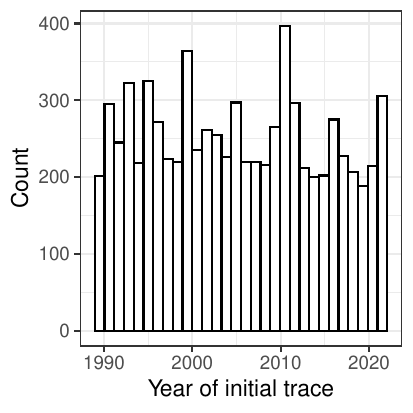}
	\end{minipage}
	\hfill
	\begin{minipage}{0.48\textwidth}
		\centering
		\includegraphics[width=\linewidth]{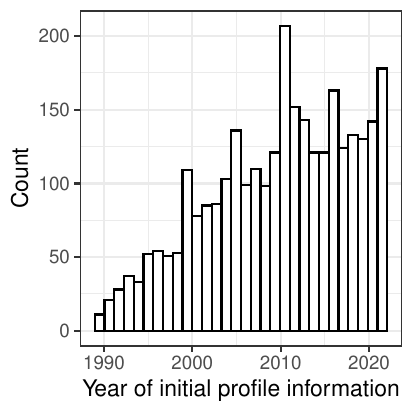}
	\end{minipage}
	\caption{Comparison of the initial year where some information is available on an individual depending on whether it is the first administrative trace (left) or the first information on the digital profile (right). 
	}
	\label{fig:period_profile}
\end{figure}

Although the initial population is exhaustively defined, many individuals are not present on LinkedIn. The proportion of individuals with a LinkedIn match varies between groups, from 25\% (cprefet) to 55\% (ENA graduates); see Table \ref{tab:kbl:3} in Appendix \ref{ap:descriptive} for all group proportions. Presence on LinkedIn is not independent of other quantities of interest. For example, Figure \ref{fig:period_profile} shows that more information is available on digital profiles for the more recent period: the year of initial trace (a proxy for the year individuals joined the population of interest) is uniformly distributed across the whole time period, but the year of initial profile information (year from which each individual lists information on LinkedIn) is strongly biased towards more recent years, both because younger civil servants are more likely to have a LinkedIn profile and because older civil servants with a LinkedIn profile may only list their more recent positions. 
This is not the only bias in the LinkedIn data: for instance, we expect individuals who move to the private sector to be more likely to have a LinkedIn profile, compared to individuals who stay in the public sector for their entire career. Taken on their own, the LinkedIn data are thus not sufficient to examine the career paths of civil servants.

We wish to understand the public-private paths at the level of the entire population, and to avoid biases that would stem from considering only individuals who are present on LinkedIn. The aim of this paper is to achieve this by augmenting the LinkedIn data with high coverage but low information data from JORF.


\subsection{Previous works on career paths of French civil servants}
\label{ss:lit}


Paths of civil servants in the private sector in France have been studied by multiple authors \citep{bouzidi_pantouflage_2010, kolopp2021pantoufler,rouban_norme_2014} and have been a source of interest in the general public. We will explore two aspects assessed as interesting in the literature but not yet described, or not described with such precision: how the structure of public-private paths has evolved since 1990 \citep{bouzidi_pantouflage_2010}; and how  patterns of these paths vary across many organizations \citep{charle_pantouflage_1987} or individual characteristics such as gender \citep{kolopp2021pantoufler}. 
Previous studies have  examined the trajectories of high civil servants in France by defining populations based on organizational belonging, but they have focused on detailed data sources only: well described career paths for a small number of people. As such, they have not explored thoroughly these questions, which require large amounts of data, and very diverse data.

\cite{rouban_inspection_2010} studied the finance general inspection (IGF) on the
1958-2008 period. IGF appears to have an important place in the
revolving doors phenomenon, as they identified that of the 578
inspectors present in the organization between 1958-2009, a majority (62\%) had
wandered at least once in the private sector. They identify a
discontinuity around 1990, with more individuals having attended a business school before joining ENA and then IGF. They also observed
after 2000 a growing proportion of IGF members who have had an
experience in the private sector before coming back to the public sector. 
However, this study  is subject to bias because it does not account for the fact that different individuals have been observed over different time scales. 

\cite{kolopp2021pantoufler} conducted a study on the general direction of the treasury
(dgtresor) in the 1965-2010 period. They exhibited a gendered behavior: men are more likely than women to leave for the most prestigious private industries (especially banking); on the other hand, they do not seem to show gender differences in the overall probability of leaving the public sector. They describe that differentiated practices generate unequal outcomes at different points in careers, for example with the stress on international mobility for career advancement and differences in professional networks. If we have reasons to assume these factors also play a significant role for other organizations, their strength may vary, and other factors might come into play. Even though we do not differentiate within private sector activities, we will compare behavior between men and women across the different organizations of our sample.

\subsubsection{Analyses of ENA students}\label{sss:ena_motivation}

One subpopulation has been the subject of particular attention in the literature and in the public eye: former students of 
the National School of Administration ENA. ENA was founded in 1945 to select and train high civil servants for the French state. Graduates of this school (around 80 to 100 a year) play a central role in the political and administrative life of the country, and they constitute a natural population when studying elite civil servants.

\cite{rouban_norme_2014} conducted a survey over 6 student year-groups ($N = 620$) who graduated in 1969-1970, 1989-90, and 1999-2000. He reports that, 10 years after graduating, 8.3\% of students from the first two cohorts were working in a private company, compared to 14.5\% for the two next cohorts, and 9.5\% for the last two, which indicates year-to-year variation but not a clear trend towards increasing probability of leaving for the private sector. 
The results of \citet{rouban_norme_2014} seem to show that the average time taken for an ENA graduate before leaving the public sector has drastically decreased over time, which would imply a stark increase in the exit probability. We must however note that their methodology does not allow for comparing different cohorts. Indeed, many statistics considered in this study, like average time to leave to the private sector, are subject to right-censoring; this was not accounted for or acknowledged. 
This is problematic as the observation frames are significantly different across cohorts: it was, at the time of publication, 44 years for the 1970 cohort, but only 14 years for the 2000 cohort. Our results on similar questions in Section \ref{ss:ena} leads to substantially different conclusions: we assume the difference with our results is due to our methodology properly taking into account the different lengths of observation frames. We conclude that the variation reported by \cite{rouban_norme_2014} is artefactual. The other quantitative conclusions of \citet{rouban_norme_2014} are not subject to this bias, and are coherent with our results.

\cite{bouzidi_pantouflage_2010} focus on a subgroup of ENA graduates, of the ``civil administrator'' status working in the Finance Ministry, between 1960 and 2002 ($N=438$), using hand-collected data. 
They show that, in their sample, about 40\% of individuals went at least once into the private sector in the 20 years following graduation. 
The advantage of their approach is that they used a formal duration analysis model, which immunizes against right-censoring biases. They however do not attempt to estimate whether there was an acceleration or not on their period of study, and stress that doing such inference is difficult for later cohorts because of right-censoring. They also estimate time spent in the private sector using simulations using two Weilbull distributions, one for the public sector and one for the private sector. Our work can be seen as a discretization of such a procedure that also enables missing data imputation and inclusion of covariates.

\subsection{Detailed and trace data sources}\label{sss:trace_types}

Unlike these previous analyses, we consider \emph{all} individuals who joined or were promoted within  the top French civil service at some point between 1990 and 2022. 
Some of these individuals stayed in the public sector for their whole career; others moved between the public and private sectors, possibly with multiple switches throughout their career. We wish to model, reconstruct, and understand these movements between public and private sectors. 

The core of our approach stems from the tension between data quality and quantity. In the majority of cases, it is difficult to obtain data that exhibit both high precision to document career trajectories, and high coverage. For French civil servants, we rely on two data sources with opposed strengths and weaknesses: 

\begin{itemize}
  \item \textbf{Detailed data source: LinkedIn.} As mentioned above, we scraped the professional networking website LinkedIn for profiles of French civil servants; this gives us the exact path over some or all of these individuals' careers. 
  LinkedIn is a
  social media platform oriented towards professional networking, where
  users describe their past work experience on their profile. 
  Many managers, both in the public and private sectors, have a
  profile, but not the entirety of the population. In addition, as these data are auto-biographical, individuals may choose to not include all relevant information: there are gaps in profiles, which means that there might be missing data even for individuals with a LinkedIn profile. This means that even when an individual is in this detailed data source, it can still be useful to complement the analysis with a trace data source. 
  This data source exhibits \textit{low coverage} but is \textit{very informative} on the individuals it documents.

  \item \textbf{Trace data source: JORF.} All civil servants leave traces in the Official Journal of the French Republic (JORF), an administrative record with a significant legal value that includes new pieces of regulation, executive decisions and nominations. 
  This source documents nominations for many top-level positions in public organizations. It also includes, in certain cases, information about promotions, other career changes (such as retirement) and decisions taken by authorities that might involve named civil servants. Individuals thus leave traces both when they are nominated or promoted, and when they sign important administrative documents. Importantly, key positions are systematically described in JORF, and this source can thus be used to define an exhaustive population. However, not all public positions are described, and there is typically no duration attached to these positions, which explains why it is not sufficient taken on its own. JORF was digitized\footnote{Data taken from Journal Officiel were first assembled and cleaned up by Nathann Cohen and put up on an API on the website steinertriples. Throughout the article, we rely on the amazing meta-data provided by this API, in addition to our own data cleaning procedures.} starting in 1990, and this defines our temporal frame (1990-2022). 
  
  A typical trace in JORF might be a record of the form "on date $t$, individual $i$ was nominated to public position $p$" or "on date $t$, individual $i$ signed a public document", both of which would show that $i$ worked for the public sector at date $t$. We usually do not have any information about when $i$ left their position, but we know that individuals in a certain career position should regularly leave traces of their activity in these records, although the rate at which this happens varies across positions and time. 
  This data source exhibits \textit{high coverage} but is \textit{not very informative} taken on its own, and it requires statistical elaboration for analysis. 
  
\end{itemize}


The LinkedIn data presents issues of heterogeneity, and data are missing not at random. We therefore expect any inference performed from the LinkedIn data alone to be biased; this is verified in practice in Section \ref{section:applied}. 
On the other hand, the trace data source is exhaustive: traces are left for legal purposes, and depend only on the position occupied by a civil servant. 
It is however not possible to perform inference from the trace data alone, as the model parameters are not identifiable.
Our modelling will therefore use both sources, and rely on the trace emission process to debias the inference. 
Concretely, it may be true that people on LinkedIn are more prone to go to the private sector, but if during their time in the public sector they leave traces   in the same fashion as people who are not on LinkedIn (and after the effect of all regressors of the trace emission model are excluded), then adding trace data will be very effective at debiasing. 
There could still be some confounding effect because individuals with digital profiles do not occupy the same positions in the public sector than individuals not on online, and that these positions would be systematically linked to fewer declarations in JORF. 
It is not immediately clear what could lead to such an imbalance, and since this confounding would come from an average across many positions, it is reasonable to expect that it is a most of a much smaller order of magnitude than the bias that would result by only including digital profiles in the study. 

This tension between complementary data sources could be found in other application areas. The detailed data source would typically correspond either to data collected by hand for a small number of individuals, or to data taken automatically from a source such as the professional networking website LinkedIn. The trace data source would correspond to larger administrative or digital records: for journalists, traces could be newspaper articles they signed; for researchers, traces would be academic papers; for the economic elite, traces might be having their names included in board meeting minutes. We focus on civil servants and do not explore these other potential applications, but we keep the methodological description general to facilitate its reuse.

\section{Model building with Markov switching models and inference}
\label{sec:model}

\subsection{Formalization}

We now turn to the formalization of this setting. We aim to study the career trajectory of $N$ individuals $i \in \{1, ..., N\}$ in discrete time $t \in \{1,...,T\}$. We denote the career position of individual $i$ in time $t$ as $X_{i,t} \in \mathcal{X}$, with $\mathcal{X}=\{0, 1\}$,  corresponding to the public and private sectors; it would be straightforward to extend to a known finite state space $\mathcal{X} = \{x_1, ..., x_K\}$  of size $K$. 
The ensemble of all trajectories at all time points is denoted $\textbf{X} = (X_{i,t})_{i,t}$. Observe that time here is discrete: our techniques may be generalized with continuous time, but we do not find a sufficient interest given the prohibitive cost that would incur. 
We take  $t$ to correspond  to a six-month period.
We assume that $\textbf{X}$ is partially observed, thanks to the \textit{detailed data source}: for some (but not all) values of $i$ and $t$, we observe the value of $X_{i,t}$.

We now need to connect it to information found in the \textit{trace data source}. The trace data source is described using a second set of time series, $\textbf{Y} = (Y_{i,t})_{i,t}$, which take value $Y_{i,t} \in \mathcal{Y}$ for each $i$ at time $t$. 
We take $Y_{i,t}$ to   be a binary variable: $Y_{i,t}=1$ if an administrative trace can be found for individual $i$ at time $t$ and $Y_{i,t}=0$ otherwise. It would be straightforward to extend this to other discrete or continuous, one-dimensional or multi-dimensional spaces. 
Unlike $\textbf{X}$, we assume that $\textbf{Y}$ is fully observed.

Lastly, this setting is described with covariates $\textbf{W} = (W_{i,t})_{i,t}$ which are of interest to the researchers or which are deemed important to control for, and which may vary across individuals and time. A typical example will be to include the sex of individuals, their age, or their initial diplomas, to see if career behavior differs across these characteristics. Importantly, these covariates will be used both to model the behavior of career paths $\textbf{X}$, and the relationship between paths $\textbf{X}$ and traces $\textbf{Y}$, and so may include both substantive and nuisance characteristics. We will show it is possible and important to include transformations of lagged values of $Y_{i,t}$ into $W_{i,t}$ when describing the relationship between $\textbf{X}$ and $\textbf{Y}$.




It is natural  to study discrete trajectories as Markov models, or, in the case there are missing values, as hidden Markov models (HMM). A complication is that we aim to retain a traditional strength of duration analyses, namely, the ability to flexibly include covariates whose importance can be assessed by statistical tests, as in proportional hazard models \citep{box-steffensmeier_event_2007}. We thus consider hidden Markov models with covariates, meaning that the transition probabilities of $\textbf{X}$ and emission distributions of $\textbf{Y}$ can vary across individuals and time. Such models are called Markov switching models. They produce good estimates yet remain easy to handle, and have been used across a wide range of applications such as clinical trials  \citep{wang_markov_1999}, finance  \citep{langrock_markov-switching_2017}, biometrics and environmental data \citep{wong_logistic_2001}, and shooting performance in sports \citep{sandri_markov_2020}.

\subsection{Formal presentation}\label{sss:formal}

For each individual $i$, we assume that the trajectory $(X_t^i)_t$ is a Markov chain over  with non-homogeneous transition kernel $P_{i,t} = (\gamma_{i,t}^{x,x'})_{x,x' \in \mathcal{X}}$. For each starting point $x\in\mathcal X$, we model the transition probabilities using a generalized linear model (GLM) with multinomial distribution:

\begin{equation}\label{eq:softmax}
  (\gamma_{i,t}^{x,x_1},..., \gamma_{i,t}^{x,x_K}) = \text{softmax}((W_{i,t}^T \beta^{x,x_1},..., W_{i,t}^T \beta^{x,x_K})),
\end{equation}
with $W_{i,t}$ the aforementioned vector of covariates, and $(\beta^{x,x'})_{x' \in \mathcal{X}}$ vectors of parameters. In this regression, we take $x\to x$ as the reference category and so set $\forall x\in \mathcal X, \beta^{x,x}=0$.  
Notice that we may have different parameters $\beta^{x,x'}$ for different transitions $(x,x') \in \mathcal{X} \times \mathcal{X}$, but that the parameters $\beta^{x,x'}$ do not depend on time. Time is rather encoded here as a variable in the covariate vectors $W_{i,t}$. It is straightforward to extend this to the multinomial case when the state space $\mathcal X$ is of cardinal greater than $2$. Notice that, for a state space of size $K$, there are only $K \times (K-1)$ vectors of parameters $(\beta^{x,x'})$ to model. In the specific case of career modeling, where individuals do not move at every step, it is reasonable to take the absence of change $x = x'$ as the reference category. 

For every point $x \in \mathcal{X}$, we next need to model the trace-emission process, that is the distribution of $Y_{i,t}$ conditional on $X_{i,t}=x$. We will use a parametric family with likelihood $\mathcal H$ and parameter $\lambda_{i,t}$ specified via  a GLM model:
\begin{equation}
  \lambda_{i,t} := \mathbb{E}[Y_{i,t}|X_{i,t} = x, \widetilde W_{i,t}] = g(\widetilde{W}_{i,t}^T \eta^x),
\end{equation}
with $\widetilde{W}_{i,t}$ the vector of covariates, $(\eta^x)_{x \in \mathcal{X}}$ vectors of parameters, and $g$ a link function. We stress that we use the notation $\widetilde{W}_{i,t}$ to communicate that when estimating emission probabilities, it is possible to include lagged values of $(Y_{i,t'})_{t' < t}$ inside $\widetilde{W}_{i,t}$, in addition to the standard covariates in ${W}_{i,t}$. Formally, $\widetilde{W}_{i,t} = [W_{i,t}^T, f(Y_{i,1},...,Y_{i,t-1})]^T$, with $f$ any user-specified transformation. We discuss why this auto-correlation feature is essential for our application in subsection \ref{sss:motivation_complexity}.

\begin{figure}[h]
  \centering
  \resizebox{.9\textwidth}{!}{
\begin{tikzpicture}[node distance=2cm, >=Stealth, auto]
	
		\tikzset{
			state/.style={
				circle,
				draw,
				ultra thick,
				minimum size=1.6cm,
				inner sep=0pt,
				align=center
			},
			every node/.style={font=\Large}
		}
		
		
		\node[state] (xt-1) {$X^{i}_{t-1}$};
		\node[state, right=of xt-1] (xt) {$X^{i}_{t}$};
		\node[state, right=of xt] (xt+1) {$X^{i}_{t+1}$};
		
		\node[state, below=of xt-1, yshift=-2cm] (yt-1) {$Y^{i}_{t-1}$};
		\node[state, right=of yt-1] (yt) {$Y^{i}_{t}$};
		\node[state, right=of yt] (yt+1) {$Y^{i}_{t+1}$};
		
		\draw[->, ultra thick] (xt-1) -- (xt);
		\draw[->, ultra thick] (xt) -- (xt+1);
		
		\draw[->, ultra thick] (xt-1) -- (yt-1);
		\draw[->, ultra thick] (xt) -- (yt);
		\draw[->, ultra thick] (xt+1) -- (yt+1);
		
		\node[left=1.6cm of xt-1] {Trajectory};
		\node[left=1.6cm of yt-1, yshift=3cm] {Covariates};
		\node[left=1.6cm of yt-1] {Traces};
		
		\node[above=0.9cm of xt] {\textbf{Standard HMM}};

		
		\node[state, right=11cm of xt-1] (xmt-1) {$X^{i}_{t-1}$};
		\node[state, right=of xmt-1] (xmt) {$X^{i}_{t}$};
		\node[state, right=of xmt] (xmt+1) {$X^{i}_{t+1}$};
		
		\node[state, below=of xmt-1,yshift=-2cm] (ymt-1) {$Y^{i}_{t-1}$};
		\node[state, below=of xmt,yshift=-2cm] (ymt) {$Y^{i}_{t}$};
		\node[state, below=of xmt+1,yshift=-2cm] (ymt+1) {$Y^{i}_{t+1}$};
		
		\node[state, below=of xmt-1, yshift=1cm, xshift=-2cm] (wmt-1) {$W^{i}_{t-1}$};
		\node[state, right=of wmt-1] (wmt) {$W^{i}_{t}$};
		\node[state, right=of wmt] (wmt+1) {$W^{i}_{t+1}$};
		
		\draw[->, ultra thick] (xmt-1) -- (xmt);
		\draw[->, ultra thick] (xmt) -- (xmt+1);
		
		\draw[->, ultra thick] (xmt-1) -- (ymt-1);
		\draw[->, ultra thick] (xmt) -- (ymt);
		\draw[->, ultra thick] (xmt+1) -- (ymt+1);
		
		\draw[->, ultra thick] (wmt-1) -- (xmt-1);
		\draw[->, ultra thick] (wmt-1) -- (ymt-1);
		\draw[->, ultra thick] (wmt) -- (xmt);
		\draw[->, ultra thick] (wmt) -- (ymt);
		\draw[->, ultra thick] (wmt+1) -- (xmt+1);
		\draw[->, ultra thick] (wmt+1) -- (ymt+1);
		
		\draw[->, ultra thick, bend left=10] (ymt-1) -- (ymt);
		\draw[->, ultra thick, bend left=10] (ymt) -- (ymt+1);
		\draw[->, ultra thick] (ymt-1) .. controls +(0,-1.5) and +(0,-1.5) .. (ymt+1);

		\node[above=0.9cm of xmt] {\textbf{Markov Switching Model}};
		
	\end{tikzpicture}}
\label{fig:tikz-models}
\caption{Comparison between standard HMM and Markov switching model.}
\end{figure}


A graphical presentation of these assumptions and a comparison with traditional HMM are available in Figure \ref{fig:tikz-models}. 
The likelihood is available in closed form for all parameter vectors for transitions $(\beta^{x,x'})_{x,x' \in \mathcal{X}}$:
\begin{equation}\label{eq:lik1}
  \forall x \in \mathcal{X}, \mathcal{L}(\beta^{x,.}| \textbf{X, W}) = \prod_{i=1}^N  \prod_{\substack{t=1\\ \text{s.t. } X_{i,t}=x}}^{T-1}  \prod_{x' \in \mathcal{X}} \left(\gamma_{i,t}^{x,x'}\right)^{1_{\{X_{i,t+1} = x'\}}}
\end{equation}
and for emissions $(\eta^x)_{x \in \mathcal{X}}$
\begin{equation}\label{eq:lik2}
  \forall x \in \mathcal{X}, \mathcal{L}(\eta^x|\textbf{X, Y, W}) = \prod_{i=1}^N  \prod_{\substack{t=1\\ \text{s.t. } X_{i,t}=x}}^{T} \mathcal{H}(Y_{i,t}|\lambda_{i,t}^x),
\end{equation}
where $\mathcal{H}$ is the likelihood function for the trace-emission GLM model, which will be practical for exploring the space of parameters with a custom sampler. In both equations, the second product is only over time indices $t$ such that $X_{i,t}=x$.

We purposefully introduced the Markov switching model in a general fashion, but  our career modeling application presents a useful specificity: certain values taken by $X_{i,t} \in \mathcal{X}$ are tied to certain values or a certain support of $Y_{i,t}$, \textit{deterministically}. 
Indeed, we know that if an individual is working in the private sector (denoted here as $X_{i,t}=\text{Private}$)
then there cannot be any traces in the public sector record. Formally, $X_{i,t}=\text{Private}$ implies $Y_{i,t} = 0$. 
There is no need to actively model the emission parameters $\eta^{x}$ in cases $x\in\mathcal{X}$ where we have such information, since, because of our external knowledge, the conditional distribution of $Y_{i,t}$ given $X_{i,t} = x$ is deterministic. This reduces the number of parameters to estimate and enables the seamless integration of expert knowledge. Additionally, when trying to reconstruct trajectories $\textbf{X}$ from traces $\textbf{Y}$, this will provide an unusually strong conditioning compared to usual applications of HMMs. 
This is a reason why our procedure works well in practice. As a whole, this application specificity thus facilitates both inference of parameters and reconstruction of trajectories.  

\subsection{Motivation from a complexity point of view}\label{sss:motivation_complexity}

The model we suggest is purposefully more complex than a Hidden Markov Model (HMM). Indeed, several issues arise from simply using a HMM in our application. One is substantive: a key interest is to describe how career behavior varies across individuals, groups, and time; thus we naturally need to include covariates in the models. Another is more technical and concerns traces. A key component of the problem is that it is essential that  the trace-emission process be well modeled: otherwise, errors here will be communicated to the rest of the procedure, which will result both in a poor reconstruction of $\textbf{X}$ and in biased inference of the $(\beta^{x,x'})$. 

Two problems make the inclusion of covariates and lagged values essential in modeling the trace-emission process: heterogeneity and self-excitation. 
We expect heterogeneity between individuals emitting traces, for instance because some civil servants make more important decisions and are promoted more often than their peers. 
If we do not account for this, our inference will confound the trace frequency variety and the actual behaviour of careers. Furthermore, it is expected that some traces will be tied or will lead to other traces in quick succession, a fact that can either be described as self-excitation or time heterogeneity: for example, civil servants joining a committee will lead to more announcements when the committee arrives at a position. Not accounting for that will lead to overestimating the probability of finding traces at any given time, and thus to overestimating the number of transitions across states of trajectories $\textbf{X}$. For this reason, using a simple HMM is not sufficient, and there is a need for both covariates and transformations of lagged values of traces. We show on synthetic data in Section \ref{ss:synthetic} that a misspecified HMM does lead to biased inference. Our results on real data in Section \ref{section:applied} confirm that this choice is appropriate: when we incorporate lagged values and gender, cohort, and organization covariates, we obtain better model fit (lower reconstruction loss) and find that many of the corresponding parameters are significantly different from 0.

\subsection{MCMC inference sampler}
\label{ss:mcmc}

Our framework relies on a large number of latent variables, making it natural to perform inference via data augmentation. 
Uncertainty quantification is essential: for instance, the career paths of certain individuals will remain highly uncertain, and we need this to translate into uncertainty about the corresponding model parameters.
More generally, we wish to allow for uncertainty to flow between the model components, and between the two data sources. We thus perform inference in a Bayesian setting with vague priors. We describe here our algorithm to sample from the joint posterior of all model parameters and latent variables.

We propose a tailored estimation scheme by MCMC to target the posterior, consisting of one Gibbs step for generating the missing values of $\textbf{X}$, and Metropolis-within-Gibbs steps for exploring the parameter spaces. This contrasts with a recent work from \cite{michelot_hmmtmb_2023} which proposes a faster (but non-exact) estimation using the Laplace approximation of the posterior. 
We give here a general presentation on our algorithm, but note that our implementation (available at \url{https://github.com/robinryder/career-paths-public}) is problem-specific: in our implementation, we only consider a state space of dimension 2, and perform several simplifications similar to those discussed in section \ref{sss:formal} and introduced in section \ref{section:applied}. The extension to the general case would be straightforward.

\textbf{Gibbs step for unobserved values of $X$}. 
The Gibbs step consists of simulating missing values of $\textbf{X}$ given observed values of $\textbf{X}$, traces $\textbf{Y}$, and parameters $(\beta^{x,x'})_{x,x'\in \mathcal{X}}$, $(\eta^x)_{x \in \mathcal{X}}$, and consists of two stages.
In the first stage, we compute the transition parameters $\gamma_{i,t}^{x,x'}$ for all individuals $i$, time $t$, and states $(x,x') \in \mathcal{X} \times \mathcal{X}$; and the conditional probabilities $p(Y_{i,t}|X_{i,t} = x, \widetilde W_{i,t}, \eta^x)$ provided by the expected value $\lambda_{i,t} = \mathbb{E}[Y_{i,t}|X_{i,t} = x, \widetilde W_{i,t}, \eta^x]=g(\widetilde{W}_{i,t}^T \eta^x)$, for all $i,t$ and states $x$.
In the second stage, we use the Baum-Welch algorithm. 
Our implementation is based on the R package \texttt{HMM} \citep{HMM}, for which we simply added a time subscript on parameters; note that the Baum-Welch algorithm allows for transition and emission parameters which are not constant across time and units. 

\textbf{Metropolis-within-Gibbs step for parameters}. Conditional on the full value of $\textbf{X}$, the inference of other model parameters corresponds to logistic regressions on different subsets of $\textbf{X}$. We take inspiration from \cite{marin2014bayesian} and apply the same method successively to different parameters of interest. Iterating on $x \in \mathcal{X}$, we update separately each vector of parameters, first $(\beta^{x,x_1},...,\beta^{x,x_K})$ jointly, and then $\eta^x$. 
For each update of an arbitrary parameter vector $\mu$, we proceed as follows: compute the MLE $\hat \mu$ for this given parameter using either equation \ref{eq:lik1} or \ref{eq:lik2} and compute the associated covariance matrix $\hat \Sigma$. Second, run for $M_{\text{Metropolis}}$ iterations a  Metropolis-Rosenbluth-Teller-Hastings random walk with the  proposition $\tilde \mu \sim \mathcal{N}(\mu, \tau^2 \hat \Sigma)$, with $\tau$ a hyperparameter that we set to 1 after experimentation. After $M_{\text{Metropolis}}$ iterations, keep the final value of parameter $\mu$ as the new value. In our applications, visual checks indicated that it was sufficient to take $M_{\text{Metropolis}}= 30$.

In practice, the first step (data augmentation) takes up the vast majority of the computational time.

\section{Empirical demonstrations on synthetic data}\label{ss:synthetic}

We present empirical demonstrations of multiple claims made in Section \ref{section:setting}, taking advantage of the notation and intuitions that arise from this specific case. 
We first show that the conjunction of detailed data and trace data leads to more accurate statistical inference: compared to using only detailed data, our credible intervals are substantially narrower.
We then demonstrate a limitation of the (simpler) standard HMM framework: even if we allow an HMM with heterogeneity in the trace-emitting process, a misspecified model leads to errors which propagate and to inferential biases; our approach can immunize against this problem.

\subsection{Gains in statistical accuracy}
\label{ss:accuracy}

We first demonstrate that information is gained by considering the two data sources, rather than only the detailed data. 
We simulate trajectories using the following form for parameters, with $g^{-1}$ the inverse logit transform: 
\begin{equation}
	\forall i, t, \ \gamma^{0\rightarrow1}_{i,t} = g^{-1}(\alpha_0+\beta_0X_i) \qquad \ \gamma^{1\rightarrow0}_{i,t} = g^{-1}(\alpha_1) \qquad \lambda_{i,t} = g^{-1}(\alpha_\lambda),
\end{equation} 
with $X_i \sim \mathcal{U}([0,1])$ an individual observed covariable, that could correspond in our example to the cohort of individual $i$. Our main parameter of interest is the parameter $\beta_0$ which measures the impact of the cohort covariable on the probability for individuals to leave the public sector.

We simulated synthetic data using the parameter values $\alpha_0=-4$, $\beta_0=1$, $\alpha_1=-4$, $\alpha_\lambda=-1.5$. 
These coefficients are similar to estimates provided for the simplest models in Section \ref{ss:ena}, so this simulation experiment can be deemed realistic. 
Each trajectory is initiated at state $0$, and is of length 60. We simulated three datasets: (1) $N=200$ individuals, all observed in full in the detailed source; (2) $N=500$ individuals, all observed in full in the detailed source; (3) $N=1000$ individuals, of which only $n_{obs}=200$ are observed in the detailed source but all of which are present in the trace data. Note that in cases 1 and 2, all paths are fully observed: the inference procedure is straightforward, and we do not expect any gain from the trace data. In case 3, we get the same number of fully observed individuals as in case 1, but we also get trace data for more individuals. This synthetic data experiment allows us to quantify the amount of information we gain from the trace data.

\begingroup
\fontsize{7}{9}\selectfont
\begin{longtable}[t]{lrrrr}
\toprule
  & Case 1 & Case 2 & Case 3 & True value\\
\midrule
$N$ & 200 & 500 & 1000 & \\
$n_\text{obs}$ & 200 & 500 & 200 & \\
$\alpha_0$ & {}[-4.22,-3.51] & {}[-4.26,-3.81] & {}[-4.19,-3.8] & -4\\
$\beta_0$ & {}[-1.04,1.25] & {}[0.17,1.64] & {}[0.72,1.96] & 1\\
$\alpha_1$ & {}[-4.91,-4.04] & {}[-4.11,-3.69] & {}[-4.36,-3.81] & -4\\
$\alpha_\lambda$ & {}[-1.54,-1.41] & {}[-1.52,-1.44] & {}[-1.5,-1.44] & -1.5\\
\bottomrule
\caption{Posterior 90\% credible intervals for the synthetic data experiment of Section~\ref{ss:accuracy}.}
\end{longtable}
\label{table:accuracy}
\endgroup

We use the MCMC algorithm described in Section \ref{ss:mcmc} to target the joint posterior distribution of $(\alpha_0,\beta_0,\alpha_1,\alpha_\lambda)$. 
We report on the concentration of the posterior of coefficients, by reporting on 90\% credibility intervals in Table \ref{table:accuracy}, from which different conclusions can be drawn. 
First, compare case 3 to case 1, and recall that these two cases have the same number of detailed observations but case 3 also includes traces: the inference in case 3 is better across the board; all 90\% posterior intervals contain the true value, but the intervals in case 3 are narrower than those in case 1, by a factor between 1.6 and 2.2 depending on the parameter of interest.
Second, cases 2 and 3 can be said to have relatively similar performances, with comparable posteriors and uncertainty. However, the gain here is that while case 2 relies on observing 500 individuals, in case 3 only 200 individuals had to be observed in the detailed data source, and imprecise knowledge using traces compensated precise knowledge on a population more than twice larger. In practice, we expect there to be a constraint on how large $n_{obs}$ can be -- either because of data availability, or because it is costly to collect detailed data on many individuals. This simulation study shows that with trace data on $N=1000$ individuals and detailed data on only $n_{obs}=200$ individuals, we get similar statistical accuracy than if we had detailed data on $N=n_{obs}=500$ individuals, a gain of a factor of 2.5 in the cost of collecting the detailed data.

\subsection{Issues with standard HMM misspecification} 
\label{ss:misspec}

We now consider a situation where a standard HMM is misspecified. 
We show that inference with a standard HMM leads to the misspecification error propagating to other inferred parameters. 
We simulate trajectories using the following form for parameters
\begin{equation}
	\forall i, t, \ 
	\gamma^{0\rightarrow1}_{i,t} = g^{-1}\left(\alpha_0+\beta_0B_{i,t}\right), \ 
	\gamma^{1\rightarrow0}_{i,t} = g^{-1}(\alpha_1), \ 
	\lambda_{i,t} = g^{-1}(\alpha_\lambda + \beta_\lambda A_{i,t}(0.8)),
\end{equation}
where $B_{i,t}=\frac{t}{60}$ is a time-varying covariate, which here could correspond to period, and $A_{i,t}(0.8)$ the autocorrelation covariate as defined earlier. 
This is a realistic scenario: as mentioned above, we don't expect traces to be emitted independently. Instead, we expect some positive autocorrelation: when an individual emits a trace, we expect them to have a higher probability of emitting further traces at the next time steps.  
We simulate synthetic data with the values $\alpha_0=-4$, $\beta_0=1$, $\alpha_1=-4$, $\alpha_\lambda=-2$, $\beta_\lambda=2.5$. 
We simulate 2000 sequences of length 60; we observe detailed data for 30\% of the individuals, and trace data for all. We implement two models: one well-specified (ours), and one misspecified (standard HMM), where covariate $A_{i,t}$ is not included, i.e. $\beta_\lambda=0$.

\begingroup
\fontsize{7}{9}\selectfont

\begin{longtable}[t]{lrrr}

\toprule
  & Well-specified & Misspecified & True value\\
\midrule
$\alpha_0$ & {}[-4.28,-3.99] & {}[-3.72,-3.49] & -4\\
$\beta_0$ & {}[0.86,2.03] & {}[-0.32,0.66] & 1\\
$\alpha_1$ & {}[-4.37,-3.95] & {}[-4.09,-3.78] & -4\\
$\alpha_\lambda$ & {}[-2.04,-1.96] & {}[-1.49,-1.45] & -2\\
$\beta_\lambda$ & {}[2.39,2.61] &  & 2.5\\
\bottomrule
\caption{Posterior 90\% credible intervals for the synthetic data experiment of Section~\ref{ss:misspec}}
\end{longtable}
\label{table:misspec}
\endgroup

We sample from the joint posterior distribution of all model parameters. 
Coefficients are reported in Table \ref{table:misspec}. 
In the well-specified case, all the output credibility intervals correctly  contain the true parameter values. 
On the other hand, in the misspecified HMM case, multiple issues appear. Not only is there no estimate for the parameter $\beta_\lambda$, but the misspecification propagates: the posterior intervals for $\beta_0$ and $\alpha_\lambda$ are far from including the true values.
Such problems will arise for time-varying covariates when an autocorrelation behavior is not taken into account in the modeling phase. 
The same point is also shown on real data in Section \ref{sss:evaluation}.

\section{Public-Private paths in French elite civil servants}\label{section:applied}

We now return to our problem of interest and apply  this method to public-private paths of French civil servants. 
We first focus on the population of ENA graduates in Section \ref{ss:ena}, then consider  all top civil servants in Section \ref{ss:orgs}. 


\subsection{Specificities in the implementation}\label{sss:specificities}

\textbf{Indices and span of the analysis}. We consider the trajectories of individuals through time $t=1, ..., T$, where an increment by one unit corresponds to 6 months: $t=1$ corresponds to January-June 1990, and $T = 68$ corresponds to July-December 2022. Each individual $i$ is modeled on an interval $[T_{min}^i,T_{\max}^i]$. We let \(T_{\min}^i\) be their first membership-defining trace (see Appendix \ref{ap:data}), which is typically the beginning of their career in the civil service. If the individual has a trace that states they retired, we use that date for \(T_{\max}^i\); else we let \(T_{\max}^i=T\), where \(T\) corresponds to our last observation time (July-December 2022). Individuals are thus modeled on different spans, a fact that is possible in the general framework but makes notation more cumbersome.

\textbf{Configuration for autoregressive traces}. We code three variables to describe the autoregressive aspect of traces. Two are given by $A_{i,t}(\varphi)= ({\sum_{0<s\leq t} \varphi^{s} Y_{t-s}^i})/({\sum_{0<s\leq t} \varphi^s})$, where we set $\varphi = 1$ to prioritize long-term information and $\varphi = 0.8$ for medium-term information. $A_{i,t}(1)$ and $A_{i,t}(0.8)$ take values between $0$ and $1$. For short-term information, we create a transformation of the time lapsed since the last trace
\begin{equation}
  L_{i,t} = \sqrt{\max \{p\ | \ \forall s<p, Y_{t-s}=0 \ \text{and} \ Y_{t-p}=1\}-1}.
\end{equation}
These three variables provide a low-dimensional and efficient encoding of past traces and tackle the issue of the varying size of past values, but many other encodings could be motivated since these are not our parameters of interest.
We used the cross-validation reconstruction loss criterion introduced in section \ref{sss:evaluation} to evaluate different possible values for $\varphi$ (we tried $\varphi = 0.6, 0.7, 0.8, 0.9$) and found that $\varphi=0.8$ minimized the loss. 

\textbf{Handling of boundary effects}. An issue is that when $t$ is low, there are not many past trace variables. This becomes problematic because a specificity of our setting is that we use traces both to define the population (as the starting positions we are interested in are systematically described) and as a complementary source (for all traces). This means that by construction, $Y_{i,T^i_{\min}} = 1$, which needs to be taken into account. As a solution, we ignore this first trace (setting the value to zero), and include specific covariates to handle points before any traces were found: $\textbf{1}_{\{A_{t,i}(1)=0\}}$, and $L_{t,i}\times \textbf{1}_{\{A_{t,i}(1)=0\}}$. A more complex handling is required when including time-sensitive coefficients, and we include instead an interaction between B-splines for age and the indicator $\textbf{1}_{\{A_{t,i}(1)=0\}}$ (see below, models 3, 4, 5 in Section \ref{ss:study1_models}).

\textbf{Prior distributions and convergence}. Results presented throughout are with non-informative, flat (improper) prior distributions. Other tests were performed with low-informative priors, e.g. $\mathcal{N}(0, 5^2)$ for model 3 in Section \ref{ss:study1_models}, with unchanged results. Contrary to fully unsupervised methods where the dependency on the prior choice could be higher, our setting only exhibits a small dependency, as points taken from the detailed data source are not simulated and thus reduce the relative importance played by the prior in the posterior distribution. We run our Metropolis-within-Gibbs sampler for  $5\,000$ Gibbs iterations with $M_{\text{Metropolis}}=30$ steps inside each iteration; we discard the first 200 as a burn-in period. Visual checks show that this is sufficient to guarantee convergence and the MCMC sampler mixes well. For all analyses, the ESS of all parameters is above 200, and the $\hat R$ is below $1.006$, indicating that the Markov chain has converged and has appropriately explored the posterior distribution. 
We repeated some analyses with $50\,000$ Gibbs iterations; these longer runs yielded similar results, confirming that the MCMC samplers have run for enough iterations.

\subsection{Subpopulation analysis: ENA graduates and temporary exit}\label{ss:ena}

We draw inspiration from the surveys of \cite{rouban_norme_2014} and \cite{bouzidi_pantouflage_2010} and first focus on civil servants who graduated from ENA. Our methodology provides two improvements. First, thanks to our methodology, we can broaden the scope and analyse all students who graduated within our time frame, including the 45\% do not have a LinkedIn profile, as well as the individuals who have a partially incomplete LinkedIn profile.
In addition, since our modeling  allows individuals to leave for the private sector and then return to the public sector, we can measure more precisely the public exit probability, and distinguish between individuals leaving the public sector permanently or temporarily.

We consider students admitted between 1990 and 2019, before the school was replaced by the National Institute for Public Service (INSP) in 2021. 
We only keep track of individuals who graduated from ENA and whose first position was  working for the state, which we operationalize by only keeping individuals for which at least one trace can be found in the three years following their graduation. The number of students identified using this method by year of admission can be found in Appendix \ref{ap:descriptive} and forms an exhaustive total of $N = 2715$. 
To our knowledge, the most substantive existing analysis of ENA graduate paths is that of \cite{rouban_norme_2014}, who focused on 6 year groups ($N=620$): our framework allows us to study a substantially larger population.

\subsubsection{Models}\label{ss:study1_models}

We implement four initial configurations which are described here from simplest to most complex. Models 1 and 2 are introduced for pedagogical purposes. \textbf{Model 1} is a standard hidden Markov model: the covariate vector $W_{i,t}$ only includes the intercept. \textbf{Model 2} is a Markov switching model for which only the autocorrelation covariates described in section \ref{sss:specificities} are included for the estimation of $\lambda$. 

Models 3, 4 (and 5) include time-specific interest covariates and notably age. Age is encoded as time elapsed since the individual was admitted to ENA, on a range from 0 to 1 (with 1 corresponding to the entire time frame, 32 years). Age is modeled in a semi-parametric fashion, using B-splines of degree 2, with 5 degrees of freedom including an intercept, and boundary knots in 0 and 1, leading to $g = 1, ..., 5$ covariates $S_{i,t}^g$ in the linear model (with the R package \texttt{splines}). We do this in order to saturate the model on the age dimension, so we can safely identify an effect related to cohort or period effects. Using splines works better than multiple powers of age: under this parameterization, parameters are less correlated, so the sampler mixes faster. As the spline basis already includes an intercept, no general intercept is included. 
Recall from our discussion in Section \ref{sss:specificities} that we have to use a more complex scheme to handle boundary effects for $\lambda$, and interact these five covariates with $\mathbb{I}_{A(1)=0}$ for all three models. More specifically, \textbf{Model 3} adds a time-varying component depending on the age of individuals for $\gamma_0$ and $\gamma_1$. \textbf{Model 4} adds a cohort effect for the estimation of \(\gamma^0\), \(\gamma^1\), and \(\lambda\). Cohort is encoded as the time at which the individual was admitted to ENA, ranging between 0 and 1.

Finally, we examine the dependence between public-private paths and electoral cycles. We implement an additional \textbf{Model 5}, which is similar to Model 4, except for the fact that instead of a cohort effect, we estimate a period effect using a dummy variable $P_{i,t}$ with value $1$ if $t$ or $t+1$ is an election time point, and $0$ else. Election time points correspond to Jan-Jun 1995, 2002, 2007, 2012, 2017, 2022 (French presidential elections). This dummy is shifted by one unit (so $t+1$ or $t+2$) for $\lambda$. 

\subsubsection{Analysis of Model Accuracy}\label{sss:evaluation}

We first analyse the output for models 1, 2, and 3. We examine both the posterior of parameters for simpler models and the quality of data augmentation by performing analyses where detailed data source information is left out for some observed individuals. 

\textbf{Parameters posterior}. The simplicity of models 1 and 2 enables full reporting on the posterior of their coefficients, which we do in Table \ref{tab:mod12} with their 0.05, 0.50 and 0.95 quantiles. 

\begingroup\fontsize{7}{9}\selectfont

\begin{longtable}[t]{lrrrrrr}
\caption{Coefficients for Models 1 and 2.}\\
\toprule
  & $M_1$ & $M_1$ & $M_1$ & $M_2$ & $M_2$ & $M_2$\\
\midrule
quantile & q05 & q50 & q95 & q05 & q50 & q95\\
$\gamma_0$, (Intercept) & -4.233 & -4.171 & -4.111 & -4.621 & -4.542 & -4.465\\
$\gamma_1$, (Intercept) & -4.054 & -3.938 & -3.830 & -3.667 & -3.525 & -3.391\\
$\lambda$, (Intercept) & -1.510 & -1.492 & -1.474 & -1.853 & -1.786 & -1.718\\
$\lambda$, $L$ &  &  &  & -0.16 & -0.134 & -0.108\\
$\lambda$, $A(1)$ &  &  &  & -0.133 & 0.088 & 0.315\\
$\lambda$, $A(0.8)$ &  &  &  & 2.027 & 2.249 & 2.467\\
$\lambda$, $I(A(1)=0)$ &  &  &  & 0.102 & 0.191 & 0.279\\
$\lambda$, $I(A(1)=0):L$ &  &  &  & -0.173 & -0.137 & -0.101\\
\bottomrule
\end{longtable}
\endgroup{}

Comparison between models 1 and 2 shows why there is a need for autoregressive coefficients and to handle boundary effects, as discussed in section \ref{sss:specificities}. The main effect of adding autocorrelation coefficients from Model 1 to Model 2 can be observed in the difference between their median intercept for \(\lambda\). For Model 1, the median intercept is at \(-1.492\) (\(I_{90}=[-1.51,-1.474]\)), which essentially corresponds to an average probability of 0.22 for an individual to generate a trace at any point in time, which is higher than the one obtained for Model 2, at \(-1.786\) (\(I_{90}=[-1.853,-1.718]\)), and which is more in line with what could be expected. All credible regions for autocorrelation coefficients except \(A(1)\) do not contain zero despite that they are quite correlated by construction. Finally, the coefficient responsible for handling the boundary effect $\mathbb{I}_{A(1)=0}$ is very different from zero and negative, showing indeed that it is needed. Configurations without it also have very different values for all other coefficients.

\textbf{Evaluation by reconstruction loss}. 
We repeat the inference under models 1, 2, 3 but with some data withheld. In this experiment, we exclude from the data 20\% of individuals in the detailed data source. We can thus assess  the data augmentation quality could on points 
that are known but excluded from the analysis. We adopt a simple cross-entropy loss on the trajectory for such points 
$l_{i,t} = - (X_{i,t}^{\text{obs}} \log(X_{i,t}^{\text{pred}}) +(1-X_{i,t}^{\text{obs}}) \log(1-X_{i,t}^{\text{pred}})$), where predictions are computed as the average of observations generated in the Gibbs step over the last $300$ iterations of the chain, with probability clipping at 0.01 and 0.99, and observations are taken from the detailed data source. For Model 1, we also implement a version where no information about the traces was used at all, which we denote Model 1b. 

The average cross-entropy loss obtained for the different models is 0.39 for Model 1, 0.29 for Model 1b, 0.22 for Model 2, and 0.21 for Model 3. A first very clear conclusion is that Model 1, the simple HMM without autoregressive components, is badly suited for the task, as it performs substantially worse than all other models. 
We observe that better reconstruction is obtained by including both data sources and autoregressive components (models 2 and 3). The change in accuracy from Model 2 to Model 3 is small in size, but our interest is rather to perform hypothesis testing when including the additional age coefficients. We might obtain better reconstruction by adopting more complex models, but this would come at the cost of model interpretability and hypothesis testing, which is our main focus. 

\subsubsection{Substantive Results}

\begin{figure}[h]
\centering
\includegraphics{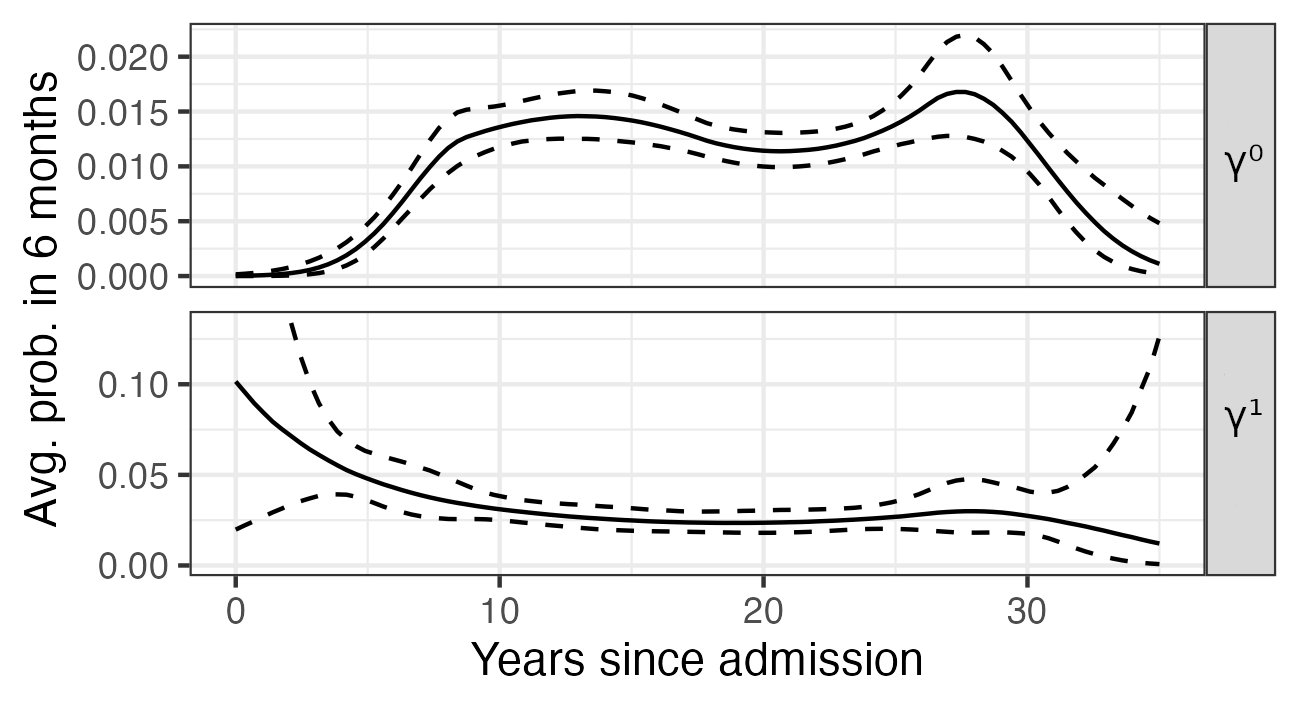}
\caption{Median marginal probabilities across time given in Model 3. $\gamma^0$ refers to the probability of leaving the public sector, and $\gamma^1$ the probability of coming back.}
\label{fig:margprob}
\end{figure}

\textbf{Dependency with age and cohort}. Figure \ref{fig:margprob} summarises information related to the probability of transitioning between states depending on age for model 3. It was computed by taking the inverse logistic transform of the sum of the intercept with the age spline coefficients. 
We first observe, that, around 0 years after admission, the probability of leaving for the private sector $\gamma^0$ is very low; this was to be expected, since individuals who never leave any trace (e.g. because they immediately resigned from ENA) are not included in the dataset.
The probability $\gamma^1$ that an individual moves back from the private to the public sector is very high, but note that this concerns a very slim proportion of the population (less than $0.7\%$ up to three years). 
The median probability $\gamma^0$ then increases to $0.013$ where it remains more or less constant up to 25 years after admission. A stabilization occurs for $\gamma^1$ at around double that value, at $0.025$. This is interesting as we observe there isn't an important variation in the probability of leaving across the career, and the hypothesis of an overall geometric distribution for leaving and coming back is quite realistic. We interpret the change of behavior 30 years after admission as artefactual, as almost no one in the population has had the time to arrive at this stage, which can also be seen with the widening of uncertainty.

\begin{figure}[h]
  \centering
  \includegraphics{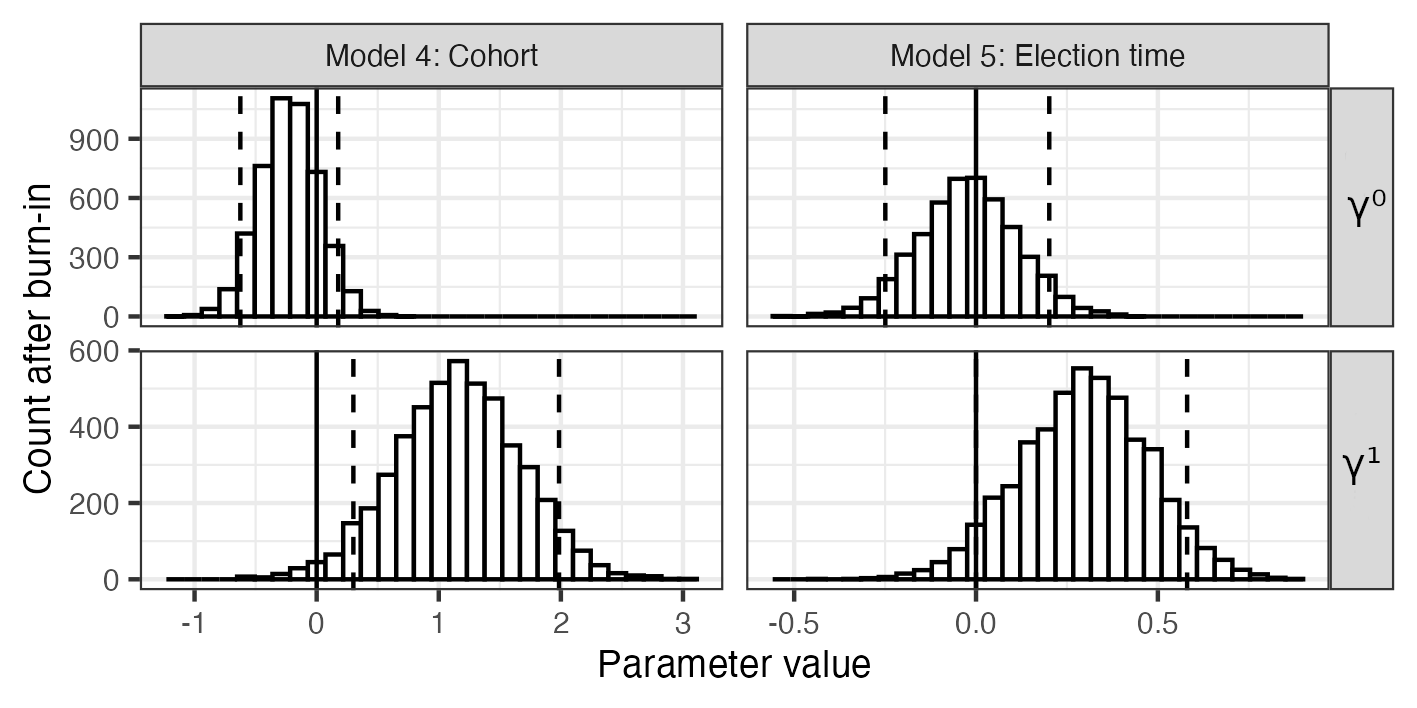}
  \caption{Posterior distribution of selected parameters for models 4 and 5. Left: effect of the cohort $\beta_{\text{cohort}}^i$. 
  Right: effect of the indicator that it is an election year $\beta_{\text{year}}^i$. Top: Posterior of the parameter $\beta^0_{\text{coeff}}$ measuring the impact on the probability of leaving the public sector $\gamma^0$. Bottom: Posterior of the parameter $\beta^1_{\text{coeff}}$ measuring the impact on the probability of returning $\gamma^1$.}
  \label{fig:post-mod45}
\end{figure}

We now turn to our main quantities of interest, cohort-related coefficients, which are
presented in the left pane of Figure \ref{fig:post-mod45}. The view is quite contrasted between
$\gamma^0$ and $\gamma^1$. On the one hand, it isn't possible to conclude
that the probability of leaving for the private sector $\gamma^0$ has significantly
changed across the period, as the credible interval well includes zero
(\(\text{med} = -0.226, I_{90} = [-0.625, 0.176]\)). However, it is at least
possible to state strongly that this coefficient has not increased across time,
and has quite possibly decreased. On the other hand, we 
can state with fair confidence that $\gamma^1$ has increased over time, 
with $\text{med}=1.16$,  (\(I_{90}=[0.301,1.98]\)). 
We thus conclude that on the one hand, the probability of leaving the public
sector has not increased in the period, and may even have decreased; and that
the probability of coming back has increased in the period, contrary
to the conclusions of \citet{rouban_norme_2014}. 

\textbf{Dependency with the electoral process}. The posterior for the election time effects are reported in the right pane 
of Figure \ref{fig:post-mod45}. We observe that the overall size of potential effects
is far smaller than for period effects, and so should be interpreted cautiously.
On the one hand, we conclude that elections have no impact on  $\gamma^0$, as the credible interval is very well centered on
zero ($\text{med} = -0.0212, I_{90} = [-0.249, 0.201]$). On the other hand, however, despite important uncertainty, there seems
to be a small positive effect of election time on 
$\gamma^1$ ($\text{med} = 0.302, I_{90} = [0.000, 0.302]$). In summary, 
electoral shifts probably do not influence the probability of high civil servants
to leave for the private sector, but it might lead  to a higher probability of 
coming back for those who had left.

\subsection{Public-private paths across the state}\label{ss:orgs}

We now move to an even larger population, and consider all top French civil servants ($N=5479$ individuals).
Our method and data allow us to describe the heterogeneity of public-private paths across the State and how they interact with gender. 

Many social scientists have early insisted on the usefulness of adopting a comparative point of view when describing civil servants populations and public-private paths \citep{charle_pantouflage_1987, kolopp2021pantoufler}. 
We attempt to build upon their results and propose a complete and
systematic comparison across the different organizations we mentioned.
This has value in the sense that it can qualify statements regarding the
most prestigious organizations that are represented in these studies:
are they really that special compared to other powerful services? For
example, are the Finance Inspection and Treasury that much more affected
by revolving doors compared to other services in the Ministry of Economics and Finance, or to
other inspections? We can also test statements linked to individual
properties: do men have a higher probability to leave for the private
sector, and then come back? 

\subsubsection{Models}\label{sss:study2_models}

We encode organizational membership using a set of \(D\) dichotomous
variables \(M_{t,1}^i,...,M_{t,D}^i\) corresponding to our
\(D = 11\) selected organizations, that we will include in the
covariate matrices \(W_t^i\). Contrary to the previous section, we need to
account for the fact that individuals can be affiliated to multiple
organizations during their professional career. To remain consistent
with our choice to focus on the main organizational identity of
individuals, we impose that someone, at any given time, may only be
linked to one organization in our set of \(D\) organizations. This means
we impose \(\sum_{d=1}^D M_{t,d}^i = 1\). If an individual is
observed to become part of organization \(d_1\) at time \(t_1\) and then
of organization \(d_2\) at time \(t_2\), we will have
\(M_{t,{d_1}} = 1\) for \(t_1 \leq t < t_2\), and
\(M_{t,d_2}=1\) for \(t \geq t_2\). This is a simplification, which has the advantage
of enabling an analysis that is closest to what we could obtain by
performing completely separate analyses for the different public
organizations, even if it could be altered to study how having been the
member of a given organization can affect behavior once being a member
of another one included in the study. 

We implement two configurations. \textbf{Model 1} is a Markov switching model for which we only include an intercept for each group (no general intercept, and no reference organization), in addition to the autoregressive covariates for traces. The autoregressive covariates are interacted with the dummy of every organization. This is our reference model. \textbf{Model 2} adds to the reference model a dichotomous variable encoding gender (\( G^i=1\  \) if \(i\) is a woman), for both \(\gamma^0, \gamma^1\) and \(\lambda\).

\subsubsection{Results}\label{sss:study2_results}

\textbf{Reconstruction loss.} The reconstruction loss obtained by the two models in this study is of $0.279$ and $0.265$, which is higher than for ENA graduates. This can be explained because the inference task is more complex given the higher heterogeneity of the population, and is in line with what could be expected given the nature of the application.

\begin{figure}[h]
  \centering
  \includegraphics{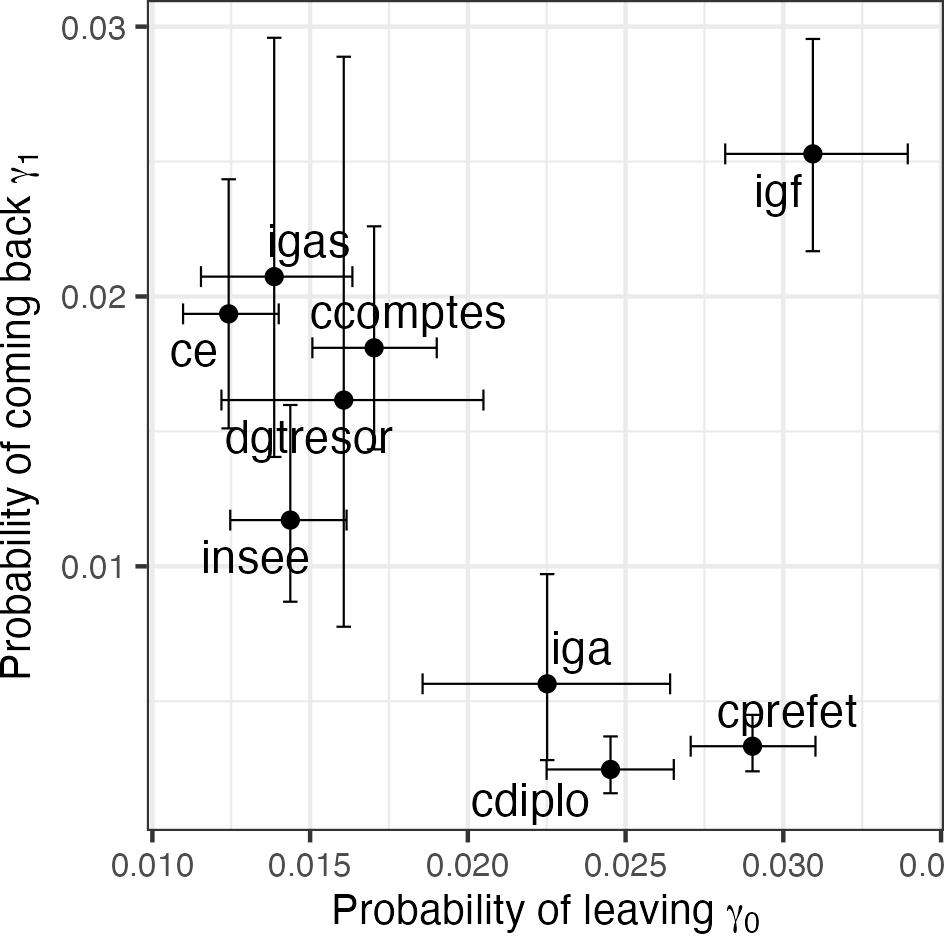}
  \caption{Median and 90\% credible interval of coefficients  $\gamma^0_\text{group}$, $\gamma^1_\text{group}$ for individuals belonging to groups defined in subsection \ref{sss:popdef}. Coefficients may be interpreted as average exit ($\gamma^0$) or return ($\gamma^1$) probabilities, yet it does not mean the model predicts transitions to happen with constant probability.}
  \label{fig:gamma-mod1}
\end{figure}

\textbf{Comparison across organizations.} We report the 5\%, 50\%, and 95\% quantiles of the posterior for each of these models. Median estimates of coefficients for Model 1 are reported in Figure \ref{fig:gamma-mod1}, alongside their 90\% credible interval, expressed
in raw probability. Bear in mind that these are average coefficients that
do not correspond to local probabilities individuals may have to leave
at any given point in time, which are modulated by the observed traces and
the local $\lambda$'s. As coefficients for dgfip stand out from the rest in both estimate and uncertainty, we do not report them in the Figure but do so in the text below.

We first rank organizations depending on the probability that
members leave for the private sector. The organizations with the highest departure probabilities are
the General Inspection of Finances
(\(\gamma^0_{\text{igf}} = 0.031\)) and Prefects group (\(\gamma^0_{\text{cprefet}}=0.029\)).
In an intermediary situation, we find the Diplomats group (\(\gamma^0_{\text{cdiplo}} = 0.022\)), and 
the General Inspection of Administration 
(\(\gamma^0_{\text{iga}} = 0.020\))
At the lower hand of the spectrum, we find the statistical office (\(\gamma^0_{\text{insee}}=0.016\)), the Cour des Comptes (\(\gamma^0_{\text{ccomptes}} = 0.015\)),
Treasury (\(\gamma^0_{\text{dgtresor}} = 0.014\)), the General Inspection for Social Affairs
(\(\gamma^0_{\text{igas}}=0.013\)), and the Conseil d'État (\(\gamma^0_{\text{ce}}=0.011\)). Not 
shown in the table, the General Direction for Public Finances has the lowest probability
(\(\gamma^0_{\text{dgfip}}=0.006, I_{90} = [0.004,0.009]\)).
This last very low coefficient means that there are almost no departures
from this direction, and when that is the case, civil servants move first to
another service included in this study. The overall ranking should be
intuitive to the reader accustomed to the question -- the model rightly
identifies that individuals in the Treasury have a higher probability of
leaving for the private sector than individuals in the Inspection for
Social Affairs. 
A seemingly counterintuitive result is the position of igf and ccomptes: these two groups
have a similar function (financial auditing of the state), but have very different positions
in Figure \ref{fig:gamma-mod1}; this observation is in fact aligned with an effect
identified in the survey from \cite{charle_pantouflage_1987}.

We observe interesting contrasts in the probabilities that individuals come
back working in public administration after being in the private sector. Inspection of Finances again stands out, with (\(\gamma^1_{\text{igf}}=0.024\)). Next, the Conseil d'État (\(\gamma^1_{\text{ce}} = 0.020\)), Inspection of 
social affairs (\(\gamma^1_{\text{igas}} = 0.019\)), Treasury (\(\gamma^1_{\text{dgtresor}}=0.018\)),
and Cour des comptes (\(\gamma^1_{\text{dgtresor}}=0.016\)) appear in an intermediate situation.
At the lower end of the spectrum, we observe the statistical office 
(\(\gamma^1_{\text{insee}} = 0.006\)), inspection of administration 
(\(\gamma^1_{\text{iga}} = 0.004\)), prefects (\(\gamma^1_{\text{cprefets}} = 0.003\)), and diplomats (\(\gamma^1_{\text{diplo}} = 0.002\)). 

We thus conclude that the two rankings are not convergent and would even be negatively
correlated if we excluded igf. However, we find it more accurate to describe this 
as forming three groups. First, igf constitutes a group on its own, standing out from the 
rest as having both high probabilities for leaving and coming back.
Classical results claiming that moving into the private sector is
part of the career advancement process hold well, and we further prove
that this is quite distinctive to this organization. We identify a second group
of organizations where the probability of leaving is relatively high, but the probability of 
coming back is rather low: this concerns the statistical office, the inspection of 
administration, diplomats and préfets. The situation is unsurprising for the first 
two, which are known to be closer to the private sector than other 
organizations. For préfets and diplomats, this might be because we 
study individuals who are simultaneously at the top of certain services but without 
as high position security as in other organizations (such as Conseil d'État,
where people can keep their status): people (involuntarily) leaving could find
themselves unable to come back. Finally, we find a group of organizations 
where the probability to leave is comparatively low and the probability of coming back high: Conseil 
d'État, Inspection of Social Affairs, Cour des Comptes. One could also place Treasury
in this group, but parameter uncertainty (especially on $\gamma^1$) requires caution.
This concerns organizations for which members acquire life-long status that can 
help them come back after leaving; and also for which clear career advancement
schemes exist in the public sector.

\begin{figure}[h!]
	\centering
	\includegraphics{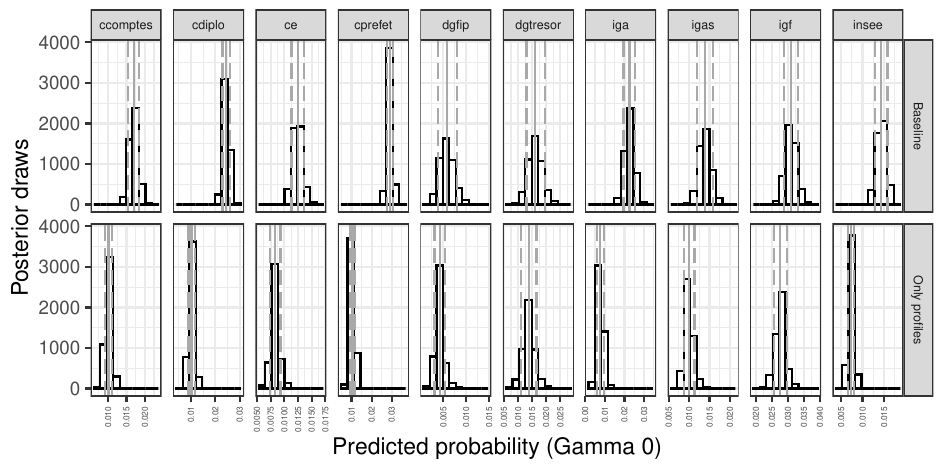}
	\caption{Comparison of results between initial study on various groups (top) and study where only individuals with a LinkedIn profile are included (bottom). Coefficients are shown in average probability scale. The posterior distributions are very different, indicating that the results on the bottom row are biased.}
	\label{fig:comp_s2_g0}
\end{figure}

\textbf{Usefulness of combining the two data sources}
We repeated the analysis using only the LinkedIn data, and show in Figure \ref{fig:comp_s2_g0} the posterior distribution of $\gamma^0$ per organization. Notice that for many groups, the posterior distribution is very different, which shows that using only the LinkedIn data leads to bias. Similar effects were observed for other parameters of interest. 
It is not possible to perform inference using only the trace data: the model parameters are not identifiable, and the MCMC algorithm therefore does not converge. 

\textbf{Interaction with gender}
We now consider the effect of the gender covariate. 
Figure 6 displays the posterior distribution of the odds ratio of $\gamma^0$ for women versus men. A distribution shifted to the right indicates that women have a higher probability of leaving for the private sector than men, and vice-versa when shifted to the left. A distribution centered
on 1 indicates no significant gender effect.

\begin{figure}[h]
\centering
\includegraphics[scale=0.8]{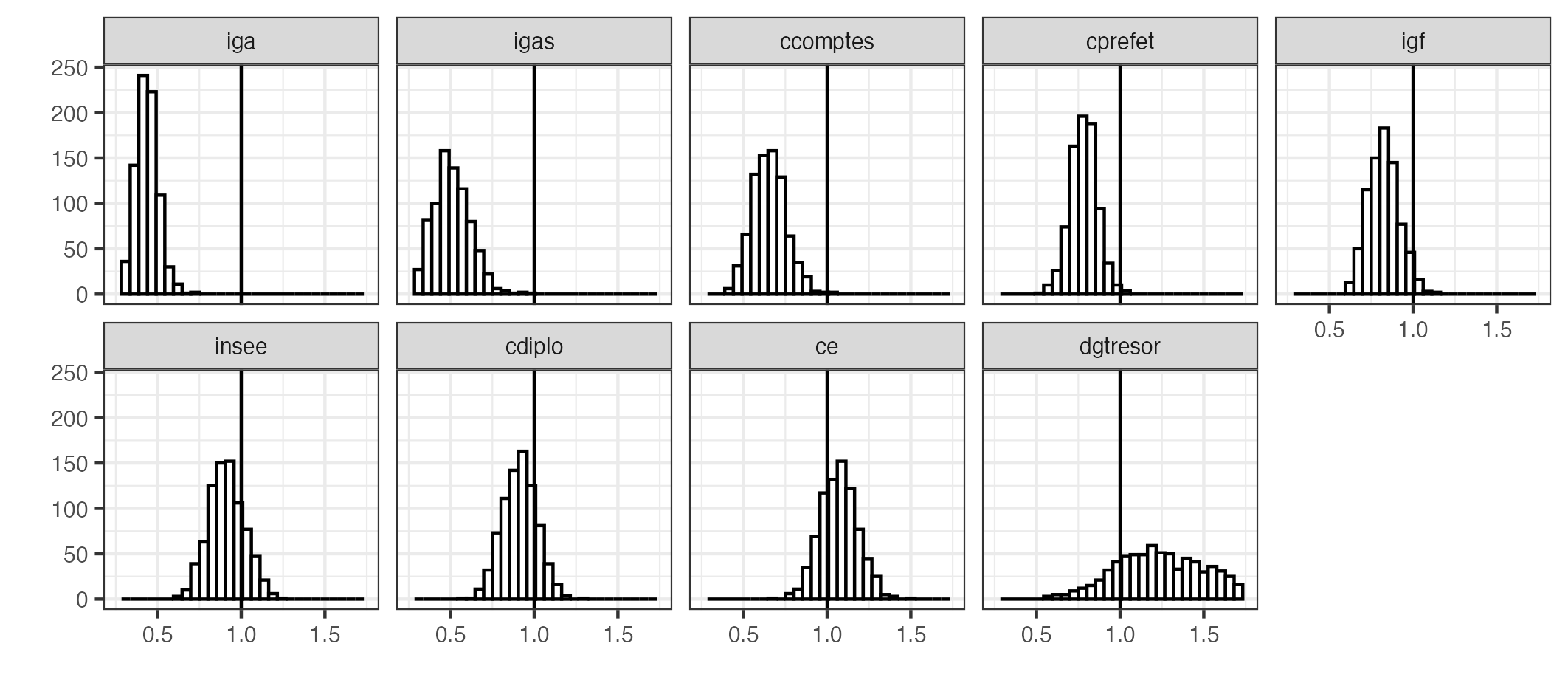}
\caption{Posterior distribution of $\frac{\gamma^0_{\text{group}, W}/ (1-\gamma^0_{\text{group}, W})}{\gamma^0_{\text{group}, M}/(1-\gamma^0_{\text{group}, M})}$, the odds-ratio of the differential of exit probabilities between women and men expressed, for each of the nine organizations defined in subsection \ref{sss:popdef}. When the posterior is concentrated on the left-hand side, we can conclude that women in that group have a lower probability than men of exiting the public sector.} 
\end{figure}

Not all organizations have the same profile of differentiated practices between 
men and women, but a broad conclusion can nonetheless be inferred: women have a lower probability than men to leave for the private sector. For certain organizations, the difference is strong and identifiable: Cour des Comptes, Inspection of Administration, Inspection of Social Affairs (odds ratios below 0.7). The difference is smaller but still observable for Inspection of Finances, Préfets, and Diplomats. The situation is undecided for Conseil État and the Statistical Office, the latter suggesting a small effect of women having a lower probability of leaving. Treasury here stands out in terms of uncertainty (as there are fewer than 200 individuals to work on and a small proportion of women), but the evidence tends to point at a higher probability of leaving for women.

\section{Discussion}\label{section:conclusion}

We have proposed a general methodology for studying career trajectories on large populations, relevant for many problems in the social sciences. This method relies on two data sources, one with high precision but low coverage, and another, consisting of imperfect traces, but with high coverage. Markov switching models with autoregressive components constitute a powerful and flexible tool that can be fruitfully used to analyze this kind of data. They can be well estimated using a custom MCMC sampler combining a Gibbs step and a Metropolis-within-Gibbs step, but remain flexible enough to test substantive research hypotheses. In addition, we showcased the issues of using misspecified HMMs using synthetic data examples and showed one could gain statistical power using our method.

Using this method, we conclude that \textbf{exit probabilities from the state have not increased on the period} 1990-2022 for  ENA graduates, contrary to what we expected before conducting this study. The data rather show that the exit probability has remained constant or has possibly decreased slightly. 
Furthermore, the probability for a former civil servants to come back to work in the public sector after leaving has significantly increased over the period. In other words, \textbf{when civil servants leave for the private sector, they now have a higher probability of coming back.} This is notably a contribution with respect to \citep{bouzidi_pantouflage_2010}, who stressed that they could not study such evolutions given the constraints of their data despite the demands of their journal referees, and showcases the advantages of studying larger groups via data augmentation.
In addition, if there seems to be an interaction between political majority changes with election years and revolving doors for this population, it seems to affect the probability of coming back from the private sector than the probability of leaving.

Across the whole elite French civil service, we find three profiles of organizations depending on the probability of leaving and the probability of coming back: one with high probability of leaving but low probability of coming back; one with low probability of leaving but high probability of coming back; and Inspection of Finances standing out from the rest with high probability of both 
leaving and coming back. Women have overall lower probability of leaving for the private sector than men, but the situation is not constant across public organizations.

Our approach has allowed us to extend the social science literature   on French civil servants and their trajectories into the private sector. An innovation in our applied studies is found in the way data were collected and cleaned. Our philosophy was to automate as much work as possible, for example in the name-matching phase, which enabled us to study a population an order of magnitude larger than what is usually done in traditional elite research (prosopography studies rarely exceed a few hundred individuals). However, we were not able to automate everything away, like public-private organization name classification, and the main limitation would again become human time if a researcher attempted to scale up this work by another order of magnitude. Our MCMC sampling procedure is more costly than approximate calculations: calculations remained very manageable in our study, but  could also become an issue when trying to scale up this approach. Enabling a better scaling of the procedure could enable to study of a larger population, and at a finer scale than based on six-month increments. 

\section*{Data and code availability}
The scripts and original data are available online at \url{https://github.com/robinryder/career-paths-public}. The MCMC output for all models mentioned in the paper is available online at \url{https://zenodo.org/doi/10.5281/zenodo.10365276}.

\section*{Acknowledgments}

The authors are grateful to six anonymous reviewers, an associate editor, and an editor for their very helpful comments on previous versions of this manuscript. 
RJR received funding from the European Union under the GA 101071601,
through the 2023-2029 ERC Synergy grant OCEAN.

\bibliographystyle{chicago}
\bibliography{refs}

@article{kolopp2021pantoufler,
	author = {Kolopp, Sarah},
	journal = {Soci{\'e}t{\'e}s contemporaines},
	number = {4},
	pages = {71--98},
	publisher = {Cairn},
	title = {Pantoufler, une affaire d'hommes? Les {\'e}narques, l'administration financi{\`e}re et la banque (1965-2000)},
	volume = {120},
	year = {2021}}

@article{seabrooke2021revolving,
	author = {Seabrooke, Leonard and Tsingou, Eleni},
	journal = {Global Networks},
	number = {2},
	pages = {294--319},
	publisher = {Wiley Online Library},
	title = {Revolving doors in international financial governance},
	volume = {21},
	year = {2021}}

@article{buhlmann2024career,
	author = {B{\"u}hlmann, Felix and Ellersgaard, Christoph Houman and Larsen, Anton Grau and Lunding, Jacob Aagaard},
	journal = {Global Networks},
	number = {1},
	pages = {e12430},
	publisher = {Wiley Online Library},
	title = {How Career Hubs Shape the Global Corporate Elite},
	volume = {24},
	year = {2024}}

@article{blanes2012revolving,
	author = {Blanes i Vidal, Jordi and Draca, Mirko and Fons-Rosen, Christian},
	journal = {American Economic Review},
	number = {7},
	pages = {3731--3748},
	publisher = {American Economic Association},
	title = {Revolving door lobbyists},
	volume = {102},
	year = {2012}}

@article{li2018journey,
	author = {Li, Pingwei and De Bosscher, Veerle and Weissensteiner, Juanita R},
	journal = {International Journal of Performance Analysis in Sport},
	number = {6},
	pages = {961--972},
	publisher = {Taylor \& Francis},
	title = {{The Journey to Elite Success: a Thirty-year Longitudinal Study of the Career Trajectories of Top Professional Tennis Players}},
	volume = {18},
	year = {2018}}

@article{machut2023emploi,
	author = {Machut, Antoine and Bastin, Gilles},
	journal = {Sociologie du travail},
	number = {3},
	publisher = {Association pour le d{\'e}veloppement de la sociologie du travail},
	title = {Emploi flexible, tournois {\`a} l'entr{\'e}e et probabilit{\'e} de durer dans le monde du journalisme},
	volume = {65},
	year = {2023}}

@book{angermuller2023careers,
	author = {Angermuller, Johannes and Blanchard, Philippe},
	publisher = {Springer Nature},
	title = {Careers of the Professoriate: Academic Pathways of the Linguists and Sociologists in Germany, France and the UK},
	year = {2023}}

@article{abbott1990measuring,
	author = {Abbott, Andrew and Hrycak, Alexandra},
	journal = {American Journal of Sociology},
	number = {1},
	pages = {144--185},
	publisher = {University of Chicago Press},
	title = {Measuring resemblance in sequence data: An optimal matching analysis of musicians' careers},
	volume = {96},
	year = {1990}}

@inbook{jackle2018temporal,
	author = {J{\"a}ckle, Sebastian and Kerby, Matthew},
	journal = {The Palgrave Handbook of Political Elites},
	pages = {115--133},
	publisher = {Springer},
	title = {Temporal methods in political elite studies},
	year = {2018}}

@manual{HMM,
	author = {Lin Himmelmann},
	note = {R package version 1.0.1},
	title = {HMM: Hidden Markov Models},
	url = {https://CRAN.R-project.org/package=HMM},
	year = {2022}}

@article{rouban_norme_2014,
	author = {Rouban, Luc},
	doi = {10.3917/rfap.151.0719},
	issn = {0152-7401, 1965-0620},
	journal = {Revue fran{\c c}aise d'administration publique},
	language = {fr},
	number = {3},
	pages = {719},
	shorttitle = {La norme et l'institution},
	title = {La norme et l'institution : les mutations professionnelles des {\'e}narques de 1970 a 2010},
	url = {http://www.cairn.info/revue-francaise-d-administration-publique-2014-3-page-719.htm},
	urldate = {2022-04-04},
	volume = {151-152},
	year = {2014}}

@article{daho_cabinet_2018,
	author = {Daho, Gr{\'e}gory and Gally, Natacha},
	issn = {0152-7401},
	journal = {Revue française d'administration publique},
	language = {fr},
	number = {4},
	pages = {849--874},
	title = {Le cabinet minist{\'e}riel comme espace fronti{\`e}re. {Les} collaborateurs minist{\'e}riels pass{\'e}s par la sph{\`e}re priv{\'e}e sous la pr{\'e}sidence de {Fran{\c c}ois} {Hollande}},
	url = {https://www.cairn.info/revue-francaise-d-administration-publique-2018-4-page-849.htm},
	urldate = {2022-05-23},
	volume = {168},
	year = {2018}}

@article{charle_pantouflage_1987,
	author = {Charle, Christophe},
	doi = {10.3406/ahess.1987.283438},
	journal = {Annales. Histoire, Sciences Sociales},

	number = {5},
	pages = {1115--1137},
	title = {Le pantouflage en {France} (vers 1880-vers 1980)},
	volume = {42},
	year = {1987}}

@article{dudouet_les_2010,
	author = {Dudouet, Fran{\c c}ois-Xavier and {Herv{\'e} Joly} and Joly, Herv{\'e}},
	doi = {10.3917/sopr.021.0035},
	journal = {Sociologies Pratiques},
	number = {2},
	pages = {35--47},
	title = {Les dirigeants fran{\c c}ais du cac 40 : entre {\'e}litisme scolaire et passage par l'{\'E}tat},
	volume = {21},
	year = {2010}}

@article{rouban_inspection_2010,
	author = {Rouban, Luc},
	doi = {10.3917/sopr.021.0019},
	issn = {1295-9278, 2104-3787},
	journal = {Sociologies pratiques},
	language = {fr},
	number = {2},
	pages = {19},
	shorttitle = {L'inspection g{\'e}n{\'e}rale des {Finances}, 1958-2008},
	title = {L'inspection g{\'e}n{\'e}rale des {Finances}, 1958-2008 : pantouflage et renouveau des strat{\'e}gies {\'e}litaires},
	url = {http://www.cairn.info/revue-sociologies-pratiques-2010-2-page-19.htm},
	urldate = {2023-08-30},
	volume = {21},
	year = {2010}}

@article{langrock_markov-switching_2017,
	author = {Langrock, Roland and Kneib, Thomas and Glennie, Richard and Michelot, Th{\'e}o},
	doi = {10.1007/s11222-015-9620-3},
	issn = {1573-1375},
	journal = {Statistics and Computing},
	language = {en},
	number = {1},
	pages = {259--270},
	title = {{Markov-Switching Generalized Additive Models}},
	url = {https://doi.org/10.1007/s11222-015-9620-3},
	urldate = {2023-09-07},
	volume = {27},
	year = {2017}}

@article{wang_markov_1999,
	author = {Wang, Peiming and Puterman, Martin L.},
	doi = {10.1080/02664769922098},
	issn = {0266-4763, 1360-0532},
	journal = {Journal of Applied Statistics},
	language = {en},
	number = {7},
	pages = {855--869},
	shorttitle = {Markov {Poisson} regression models for discrete time series. {Part} 1},
	title = {{Markov {Poisson} Regression Models for Discrete Time Series. {Part} 1: {Methodology}}},
	url = {http://www.tandfonline.com/doi/abs/10.1080/02664769922098},
	urldate = {2023-09-07},
	volume = {26},
	year = {1999}}

@book{marin2014bayesian,
	author = {Marin, Jean-Michel and Robert, Christian P},
	publisher = {Springer},
	title = {Bayesian Essentials with R},
	volume = {48},
	year = {2014}}

@misc{michelot_hmmtmb_2023,
	author = {Michelot, Th{\'e}o},
	language = {en},
	note = {arXiv:2211.14139},
	publisher = {arXiv},
	shorttitle = {{hmmTMB}},
	title = {{hmmTMB}: {Hidden} {Markov} models with flexible covariate effects in {R}},
	url = {http://arxiv.org/abs/2211.14139},
	year = {2022}}

@article{box-steffensmeier_event_2007,
	author = {Box-Steffensmeier, Janet M. and De Boef, Suzanna and Joyce, Kyle A.},
	doi = {10.1093/pan/mpm013},
	issn = {1047-1987, 1476-4989},
	journal = {Political Analysis},
	language = {en},
	number = {3},
	pages = {237--256},
	shorttitle = {Event {Dependence} and {Heterogeneity} in {Duration} {Models}},
	title = {Event {Dependence} and {Heterogeneity} in {Duration} {Models}: {the} {Conditional} {Frailty} {Model}},
	url = {https://www.cambridge.org/core/product/identifier/S1047198700006495/type/journal_article},
	urldate = {2023-09-21},
	volume = {15},
	year = {2007}}

@book{france_sphere_2017,
	author = {France, Pierre and Vauchez, Antoine},
	publisher = {Les Presses de Sciences Po},
	title = {Sph{\`e}re publique, int{\'e}r{\^e}ts priv{\'e}s. {Enqu{\^e}te} sur un grand brouillage},
	url = {https://journals.openedition.org/lectures/22901},
	urldate = {2023-10-11},
	year = {2017}}

@article{wong_logistic_2001,
	author = {Wong, C. S.},
	doi = {10.1093/biomet/88.3.833},
	issn = {0006-3444, 1464-3510},
	journal = {Biometrika},
	language = {en},
	number = {3},
	pages = {833--846},
	title = {{On a Logistic Mixture Autoregressive Model}},
	url = {https://academic.oup.com/biomet/article-lookup/doi/10.1093/biomet/88.3.833},
	urldate = {2023-10-11},
	volume = {88},
	year = {2001}}

@article{sandri_markov_2020,
	author = {Sandri, Marco and Zuccolotto, Paola and Manisera, Marica},
	journal = {Journal of the Royal Statistical Society Series C: Applied Statistics},
	number = {5},
	pages = {1337--1356},
	title = {Markov switching modelling of shooting performance variability and teammate interactions in basketball},
	url = {https://academic.oup.com/jrsssc/article-abstract/69/5/1337/7058675},
	urldate = {2023-10-11},
	volume = {69},
	year = {2020}}

@article{bouzidi_pantouflage_2010,
	author = {Bouzidi, Btissam and Gary-Bobo, Robert and Kamionka, Thierry and Prieto, Ana},
	doi = {10.3917/rfe.103.0115},
	issn = {0769-0479, 1760-7388},
	journal = {Revue fran{\c c}aise d'{\'e}conomie},
	language = {fr},
	number = {3},
	pages = {115},
	shorttitle = {Le pantouflage des {\'e}narques},
	title = {Le pantouflage des {\'e}narques : une premi{\`e}re analyse statistique},
	url = {http://www.cairn.info/revue-francaise-d-economie-2010-3-page-115.htm},
	urldate = {2023-10-31},
	volume = {XXV},
	year = {2010}}

\appendix

\section{Data assemblage}\label{ap:data}

This appendix documents the various procedures used to construct the population and datasets. Subsection \ref{exact-perimeter-for-organizational-belonging} describes how the population is defined based on a thoroughly constructed \textit{subset} of administrative traces in JORF. Subsection \ref{selection-of-administrative-traces} describes how, more generally, decisions in JORF are handled to construct the trace data source. This process is distinct from the population definition task done in subsection \ref{exact-perimeter-for-organizational-belonging} and relies on a much wider array of administrative traces. Subsection \ref{linkage-with-digital-profiles} describes how linkage to the detailed data source (Linkedin) is performed. The procedure implemented to handle the risk for homonyms and false positive matchings is described in detail.  

\subsection{Exact perimeter for organizational belonging}
\label{exact-perimeter-for-organizational-belonging}

To construct the population in this article, a restrictive set of traces from JORF is used, with much more labor done by hand and back and forth in discussion with sociologists. This enabled to arrive at a reasonably grounded definition of what constitutes membership in any of the groups discussed in the article. Given the complexity of the legal organization of French Grands corps, it is reasonable that other definitions could be reached or argued for, and our choice is thus to exhaustively share here the decision-rule system that was used to define membership in a group.

We expect that a social scientist working on any specific group could motivate a finer definition of organizational membership, with further distinctions to be made within groups. Readers should thus keep in mind that our substantive contribution is essentially comparative, and that further variation ought to be discovered at a finer level. Further, readers should also note that the definitions are necessarily constrained by the nature of the archive and what it documents. If a career transition cannot legally be found in the JORF, then, it cannot be used to define populations in the manner proposed here. This is a problem for many trajectory studies, but since for elite civil servants, nomination procedures are on the highly-formalized end of the spectrum, we find that problem to be manageable. However, it is clear that the following definitions are, like the organizational boundaries they are trying to approximate, socially constructed and context-dependent. We find there is already a contribution in making the decision rule transparent so that it can be further improved or specialized for other studies. 

Concretely, we begin by surveying the Steinertriples API for the different groups selected for the study. Traces are first filtered depending on their tagged organization and their nature, using a rule-based approach summarized in Table \ref{tab:kbl:codex} (0 means filtered out, 1 means kept in). This was guided using substantive knowledge of civil servants and asking some help from sociologists. Some decisions were taken by studying specific biographies of individuals that would be included or excluded from the lists by any decision, to see if it would correspond to an intuitive definition of membership. 

\begin{table}

\caption{\label{tab:kbl:codex}Type of traces used to define membership by organization}
\centering
\begin{tabular}[t]{lrrrr}
\toprule
short\_name & affectation & nomination & intégration & titularisation\\
\midrule
cab & 0 & 1 & 0 & 0\\
ccomptes & 1 & 1 & 1 & 0\\
cdiplo & 0 & 1 & 0 & 0\\
ce & 1 & 1 & 0 & 1\\
cprefet & 0 & 1 & 0 & 1\\
csprefet & 0 & 1 & 0 & 1\\
dgbudget & 1 & 1 & 1 & 1\\
dgfip & 1 & 1 & 1 & 1\\
dgtresor & 1 & 1 & 1 & 1\\
ena & 0 & 1 & 0 & 0\\
iga & 1 & 1 & 1 & 1\\
igas & 1 & 1 & 1 & 1\\
igf & 1 & 1 & 1 & 1\\
insee & 0 & 1 & 0 & 0\\
\bottomrule
\end{tabular}
\end{table}

After this first filter, we apply specialized regex filters to precisely filter the population so it resembles as much as possible expert interpretations. 

\begin{itemize}
  \setlength\itemsep{0em}
  \item For ccomptes, traces that are not affectation, we only keep matches of (auditeur)|(auditrice)|(conseiller)|(conseillère)|(président)|(magistrat). We include the array of elite civil servants working at the Court. 
  \item For insee, we only keep traces that match (administrat)|(inspect)|(direct) and do not match (regiss)|(élève). We thus focus on members of the legal insee grand corps. 
  \item For ce, traces that are not affectation, we only keep traces matching (auditeur)| (auditrice)|(conseiller)|(conseillère)|(maître). We focus on Conseillers d'État and auditors.
  \item igas, iga, igf, for nomination traces, we only keep matches of (inspecteur) |(inspectrice). We thus focus on inspectors. 
  \item For dgtresor and dgfip, nomination traces, we only keep matches of (directeur) |(directrice)|(chef)|(administrat)|(délégué)|(attaché)|(conseiller). We also remove nomination traces matching (commission)|(expert)|(régisseur)| (suppléant). For dgfip, we also remove traces that document being a member of a cabinet. We thus mostly focus on directors and sub-directors of services and economics experts. 
  \item For ENA, we only keep traces of admitted students. 
\end{itemize}

After applying this filter, we then remove nomination traces matching (membre) |(représentant) which correspond to internal elections.

\hypertarget{selection-of-administrative-traces}{%
\subsection{Selection of administrative
traces}\label{selection-of-administrative-traces}}

For every person in our population, we query the Journal Officiel to obtain every administrative order carrying the same name, which yields a first set of traces we name \(S_1\). As many people have perfect homonyms, we have to filter this set. Most traces are tagged with one or multiple organizations, which range from small bureaus to the entire police force. To reduce this risk of perfect homonyms, we only keep traces linked to organizations significantly over-represented in \(S_1\) compared to the entire Journal Officiel. For each organization with the set \(O\) of traces linked to it, we compute:\\
\[
\frac{\mathbb P[O|S_1]}{\mathbb P[O]} = 
\frac{\# \{\text{traces linked to org} \ O\} \cap S_1}{ \sum_x \# \{\text{traces linked to org} \ x\} \cap S_1} /
\frac{\# \{\text{traces linked to org} \ O\} }{ \sum_x \# \{\text{traces linked to org} \ x\}}
\] 
which can be interpreted as a relative risk score. We only keep traces linked to organizations with a score above 2, or above 1 if they have fewer than 1000 traces, or above 0.5 if they have fewer than 100 traces. This effectively removes organizations such as the army or the police where many promotions are described for the entire force, and
constitute the main sources of perfect homonyms in \(S_1\). For traces that are not linked to any organization, we also include them if they are a ``d\'el\'egation de signature'' (meaning an authority delegates administrative power to the civil servant, a practice that remains limited in scope) or if it states an admission into a ``grand corps''. This enables a good tradeoff between minimizing the homonymy risk and keeping as much data as possible, and yields a second set of traces \(S_2\). For the first population, we then manually search for every individual for which we cannot find any trace upon graduation, and document the cases of name-changing for individuals who got married during their time as students (around 200 individuals) and add the relevant traces to the total after filtering them using the same procedure. For the second population, if a trace in \(S_2\) documents the person has had another name, we also add the relevant traces after passing them through the same filters as before (this also represented around 200 individuals, which were also all checked by hand). This yields our final set of traces \(S_3\). This set is the one used as the ``traces'' dataset throughout the article. 

\subsection{Linkage with digital profiles}\label{linkage-with-digital-profiles}


For each profile linked to an individual, we obtain a list of dated professional positions that are shared between the public and private sectors (or ``spheres'', see \cite{daho_cabinet_2018,france_sphere_2017}). We classify organization names between public and private positions, using a rule-based method, progressively adding filters to capture all public organizations. This strategy enabled us to cover manually the entire sample. Companies entirely owned by the state (e.g. La Poste, France Télévisions) are also coded as public, alongside companies with the ``EPIC'' status (e.g.~SNCF, OSEO). Contrary to \cite{daho_cabinet_2018}, companies that only have the state as a minority shareholder (e.g.~Renault, TotalEnergies) are coded as private. In addition, positions at Universities and research bodies are not coded as either public or private: it appears they correspond to associate teaching positions that do not constitute the main occupation for people in our population, and so we leave them out. This reasoning leads us to also not code implication in a political campaign, as this is often superposed to other engagements. The same goes for positions in associations, naturally including charities, but also positions in unions, which include corporate representation bodies such as Medef. Finally, to ensure homogeneity with Journal Officiel, we do not include positions in world organizations (e.g.~World Health Organization). This strategy of not coding uncertain cases is enabled by our modeling approach, as it doesn't require full trajectories, and reduces the impact of such decisions on results. Without claiming this coding scheme is perfect, these decisions only have a minor impact on the substantial research findings, and all 1488 coding filters will be made fully accessible in the appendix for re-use.

\section{Further descriptive statistics}\label{ap:descriptive}

The following tables provide descriptive statistics. Table \ref{tab:kab:1} describes the population size of ENA students included each year in the study. Table \ref{tab:kab:2} describes the population size of the different groups of elite civil servants (note that some individuals may be counted twice). Table \ref{tab:kbl:3} gives the proportion of individuals, in any of the groups that are matched to a digital profiles. Table \ref{tab:kab:4} describes the gender proportion  and the proportion of ENA graduates by organization.

\begin{table}

\caption{\label{tab:kab:1}Number of students by year of admission.}
\centering
\begin{tabular}[t]{lrlrlr}
\toprule
Year & Count & Year & Count & Year & Count\\
\midrule
1990 & 89 & 2000 & 107 & 2010 & 76\\
1991 & 97 & 2001 & 112 & 2011 & 74\\
1992 & 81 & 2002 & 135 & 2012 & 80\\
1993 & 97 & 2003 & 112 & 2013 & 77\\
1994 & 104 & 2004 & 99 & 2014 & 75\\
1995 & 95 & 2005 & 90 & 2015 & 79\\
1996 & 96 & 2006 & 83 & 2016 & 87\\
1997 & 96 & 2007 & 87 & 2017 & 85\\
1998 & 95 & 2008 & 78 & 2018 & 77\\
1999 & 98 & 2009 & 79 & 2019 & 75\\
\bottomrule
\end{tabular}
\end{table}

\begin{table}

\caption{\label{tab:kab:2}Number of individuals affiliated by group (1990-2022).}
\centering
\begin{tabular}[t]{lrlr}
\toprule
group & count & group & count\\
\midrule
insee & 1164 & igf & 455\\
cdiplo & 914 & igas & 402\\
cprefet & 838 & dgfip & 373\\
ccomptes & 629 & iga & 203\\
ce & 626 & dgtresor & 171\\
\bottomrule
\end{tabular}
\end{table}

\begin{table}

\caption{\label{tab:kbl:3}Proportion of individuals by group with a matched profile on LinkedIn.}
\centering
\begin{tabular}[t]{llll}
\toprule
group & prop & group & prop\\
\midrule
ena & 55.2 & ce & 37.9\\
dgtresor & 54.4 & insee & 33.2\\
igf & 52.3 & dgfip & 32.4\\
igas & 44.5 & cdiplo & 26.8\\
ccomptes & 41.5 & cprefet & 24.7\\
iga & 37.9 &  & \\
\bottomrule
\end{tabular}
\end{table}

\begin{table}

\caption{\label{tab:kab:4}Proportion of men and ENA graduates by organization.}
\centering
\begin{tabular}[t]{lrrlrr}
\toprule
group & prop\_men & prop\_ena & group & prop\_men & prop\_ena\\
\midrule
cprefet & 87.6 & 11.7 & dgfip & 72.4 & 11.8\\
cdiplo & 82.7 & 8.4 & iga & 71.4 & 31.0\\
dgtresor & 80.7 & 26.3 & ce & 66.3 & 34.5\\
ccomptes & 78.2 & 31.3 & insee & 64.8 & 0.3\\
igf & 75.8 & 37.8 & igas & 61.5 & 29.8\\
\bottomrule
\end{tabular}
\end{table}

\section{Additional robustness checks in the choice of $\varphi$}\label{ap:robustness}

In our implementation, we include the covariates $A(1)$ and $A(\varphi)$ with $\varphi=0.8$.
We performed additional tests with different weighting parameters $\varphi$ for the main model of interest (model 3 in Section \ref{ss:study1_models}) $\varphi = 0.6, 0.7, 0.8, 0.9$, and compared with the reconstruction loss criterion ($0.222, 0.239, 0.216, 0.229$). The best-performing model is when choosing $\varphi = 0.8$.

We note that the inference behaves in a very similar manner when setting $\varphi = 0.6, 0.7, 0.8$, but changes when setting $\varphi = 0.9$. When setting $\varphi = 0.9$, $A(\varphi)$ and $A(1)$ are very correlated (correlation  $0.93$) and redundant, and the model is not able to pick up on medium-term behavior of the trace-emission process, which was the reason why we introduced $A(\varphi), \varphi < 1$ in the first place, with some notable change in the parameter posteriors.
Based on this, and since other variables already have the purpose of tracking small-term and long-term changes, we argue that one should avoid setting $\varphi$ too high (or too low). 

For additional verification, we also constructed a model that includes multiple values of $\varphi$ ($\varphi= 0.6, 0.8$), in addition to $\varphi = 1$ (it is not possible to include all $\varphi = 0.6$ to $0.9$ because it leads to an ill-conditioned linear system). The reconstruction loss is very close to the original model ($0.216$), and the parameters inferred are also very similar, with no change in substantive conclusions.

\section{Digital appendix and data availability}\label{ap:digital}

The scripts and original data are available online at \url{https://github.com/robinryder/career-paths-public}. The MCMC output for all models mentioned in the paper is available online at \url{https://zenodo.org/doi/10.5281/zenodo.10365276}.

A note on the type of data released with this paper is in order. The released data is at an individual level with a 6-months precision frame, and is fully anonymized. Only abstract numeric identifiers are released, alongside the following information: gender, public organization group covered in the study (e.g. ENA, Conseil d'État), whether a trace in JORF was observed in a given 6-months time frame, whether the identifier is linked to a digital profile, whether the person is working in the public or private sector in a given six-months time frame according to their digital profile. All other columns of the dataframe released are computed from this basic information. This is exactly the strictly required information for the models trained and presented in the paper, and release is mandatory so that the proposed statistical techniques are transparent and replicable.

To avoid risks of de-anonymization, no other information is released or can be released to the public. For example, specific private organization names or public organization names besides the exact list studied in the paper are not released. This, alongside the time granularity of six months, ensures that the cost of de-anonymization of this study's data is lower than the cost of assembling the data from the ground up, making de-anonymization an essentially useless endeavor. The information, despite being linked to an individual identifier, is also sufficiently crude so that matching it to another dataset cannot provide any substantial advantage, making de-anonymization more than just costly, but also without benefit. For this reason, it is also not possible to share the scripts that were used to originally collect the data, especially for the digital profiles part, as sharing them would be essentially equivalent to releasing the uncensored data. Even if the whole data pipeline run at any given time relies exclusively on publicly available information, it still takes as input individual data. As such, only the statistical inference scripts are made public and no other scripts can be made public.  

Finally, despite it being very clear from the paper's framing, let us explicitly state that the statistical analysis is purely motivated from a structural, scientific, analysis of public-private paths of elite civil servants. Data was collected without any interest for any specific individual, and the sole interest was to describe trends. It is anchored in a strong statistical and social scientific framing, and is also motivated by a clear public interest question, that of the deontology of elite civil servants and the breadth of revolving doors in France put in the historical context of a 30 years time frame.

\end{document}